\documentclass[conference]{IEEEtran}
\usepackage{cite}
\usepackage{amsmath,amssymb,amsfonts}
\usepackage{algorithmic}
\usepackage{graphicx}
\usepackage{textcomp}
\usepackage{xcolor}
\usepackage[hyphens]{url}
\usepackage{fancyhdr}
\usepackage[bookmarks=true,breaklinks=true,colorlinks,citecolor=blue,linkcolor=blue,urlcolor=blue]{hyperref}

\usepackage{animate}

\def\BibTeX{{\rm B\kern-.05em{\sc i\kern-.025em b}\kern-.08em
    T\kern-.1667em\lower.7ex\hbox{E}\kern-.125emX}}

\usepackage{mathptmx} 
\usepackage{amsmath,amssymb,amsfonts}
\usepackage{graphicx}
\usepackage{textcomp}
\usepackage{xcolor}
\usepackage{color,soul}
\usepackage{booktabs}
\usepackage{tabularx}
\usepackage{balance}
\usepackage{amsbsy}
\usepackage{pifont}
\usepackage{listings}
\usepackage{xspace}
\usepackage[normalem]{ulem}

\usepackage{cleveref}
\crefformat{section}{\S#2#1#3}
\crefformat{subsection}{\S#2#1#3}
\crefformat{figure}{Fig.~#2#1#3}
\crefformat{listing}{Listing~#2#1#3}
\crefformat{equation}{Eq.~(#2#1#3)}

\definecolor{darkred}{rgb}{0.8, 0.0, 0.0}
\definecolor{darkgreen}{rgb}{0.0, 0.5, 0.0}
\definecolor{mustard}{rgb}{1.0, 0.77, 0.05}

\newcommand{\proj}{Tascade\xspace}

\author{
Marcelo Orenes-Vera, Esin Tureci, David Wentzlaff, Margaret Martonosi\\
Princeton University, Princeton, New Jersey, USA \\
\{movera, esin.tureci, wentzlaf, mrm\} @princeton.edu
}
\newcommand{\repository}{repository\cite{tascade_repo_names}\xspace}

\title{Tascade: Hardware Support for Atomic-free, Asynchronous and Efficient Reduction Trees}

\begin{document}
\maketitle
\pagestyle{plain}

\newcommand{\subf}[2]{%
  {\small\begin{tabular}[t]{@{}c@{}}{\tiny}
  #1\\#2
  \end{tabular}}%
}

\begin{abstract}

Graph search and sparse data-structure traversal workloads contain challenging irregular memory patterns on global data structures that need to be modified atomically. Distributed processing of these workloads has relied on server threads operating on their own data copies that are merged upon global synchronization. As parallelism increases within each server, the communication challenges that arose in distributed systems a decade ago are now being encountered within large manycore servers. Prior work has achieved scalability for sparse applications up to thousands of PUs on-chip, but does not scale further due to increasing communication distances and load-imbalance across PUs.

To address these challenges we propose Tascade, a hardware-software co-design that offers support for storage-efficient data-private reductions as well as asynchronous and opportunistic reduction trees. Tascade introduces an execution model along with supporting hardware design that allows coalescing of data updates regionally and merges the data from these regions through \emph{cascaded updates}. Together, Tascade innovations minimize communication and increase work balance in task-based parallelization schemes and scales up to a million PUs.

We evaluate six applications and four datasets to provide a detailed analysis of Tascade's performance, power, and traffic-reduction gains over prior work. Our parallelization of Breadth-First-Search with RMAT-26 across a million PUs---the largest of the literature---reaches over 7600 GTEPS.

\end{abstract}

\vspace{-1mm}
\section{Introduction}
\vspace{-0.5mm}

Programs whose loops can be parallelized have the underlying assumption that any interleaving of operations across different iterations preserves correctness.
Many graph and sparse applications have associative and commutative reduction operations (e.g., minimization), making them amenable to such parallelization, provided that all writes to the same data happen atomically.
This is the underlying power of models such as Bulk-Synchronous Parallelization (BSP) of graph applications as they guarantee eventual correctness even with arbitrary interleavings of all other read/write operations.

Distributed-memory MapReduce~\cite{mapreduce} or data-private reductions~\cite{openmp_privatization,upc_privatization} may avoid atomic operations by having processors operate on local copies of the reduction array that are merged upon global synchronization.
However, since pre-merge computations and the merging process in itself cannot be overlapped in such software schemes, it leads to idle PUs towards the end of the computation phase.
This under-utilization is exacerbated in the context of BSP, where each graph search epoch has a global barrier.
Moreover, these approaches incur storage overheads proportional to the number of copies, i.e., up to the number of software threads used.

With the recent rise of massively parallel AI-oriented manycores~\cite{cerebras,groq,tesla_dojo,tpu_google,nvidia_a100} and the unmet demand for systems that can accelerate graph and sparse data-structure traversal~\cite{iarpa_agile,agile_hpcwire}, we see the \textbf{opportunity} to support more efficient and scalable reduction operations within these systems.

Dalorex~\cite{dalorex} demonstrated scaling of graph and sparse applications to thousands of PUs without storage overheads by proposing a task-based execution model on top of tile-based distributed-memory manycores.
Dalorex distributes the dataset arrays in equal-sized chunks across the tile grid, such that each tile holds a chunk of the dataset which only the tile's PU can operate on, making all operations inherently atomic.

Although at first sight Dalorex's single-data-owner model appears to solve the scaling problem, it reveals two new bottlenecks when scaling beyond thousands of PUs:
(1) Since only a single tile's PU can operate on a given data, work imbalance becomes significant when the underlying dataset is skewed, limiting PU utilization as the scaling increases;
(2) As the system size increases, task invocations across PUs must travel longer distances on average, increasing network congestion.
In order to scale to millions of PUs, a new approach must aim for to achieve these \textbf{goals}: \emph{scale-invariant communication}, \emph{PU work-balance} and \emph{network-load balance}.

We present \proj, which makes strides in these three goals by implementing lightweight hardware support for scalable and storage-efficient data-private reductions as well as asynchronous and opportunistic reduction trees to merge updates across the network.
These design innovations enable \proj to scale the parallelization of graph and sparse applications to millions of PUs.

\textbf{\emph{Data-private reductions through Proxy Caches:}}
To reduce long-distance communication, \proj divides the tile grid into subgrids called \emph{proxy regions}, depicted in \cref{fig:reduction_tree}.
Each region is assigned responsibility for a copy of the reduction array (called \emph{proxy array}), equally divided among the tiles in the region.
As such, \proj relaxes the \emph{single-owner-per-data} constraint of Dalorex, and permits proxies to operate on the proxy array via cache structure called \emph{proxy cache} (P-cache).
From here on in this paper, we call the tile that owns the original reduction array chunk the \emph{owner tile} to distinguish it from the proxies.
Tasks requiring long-distance communication are first sent to the proxy owner within the sender's region.
P-caches are key components of the reduction trees as well (see below) where the write-propagation policy of P-cache determines when a merge operation is triggered (see \cref{sec:proxy_cache}).
The P-cache and proxy region sizes are software configurable to balance the trade-off between storage overhead and communication distance (elaborated in \cref{sec:proxy_cache_config} and evaluated in \cref{sec:res_proxy_region_size}).

\textbf{\emph{Reduction Tree Implementation for Merge operations:}}
Merge operations occur when the update is sent to the owner tile based on the write-propagation policy of the P-cache (also software configurable).
Upon sending updates from the proxy tile to the owner tile, additional proxies en route may capture the task as proxy tasks through a mechanism called \emph{Selective Cascading}.
This effectively results in the implementation of a reduction tree for each owner tile (which behaves as the root of the tree).
Such parallel, asynchronous and distributed reduction tree approach along with the data-private reductions through P-caches minimize communication through filtering/coalescing of regional updates; improve work balance by parallelizing data reductions over multiple tiles, and achieve network-load balancing through selective cascading.

\textbf{\emph{Hardware Support for Cascading Updates:}}
\proj introduces a hardware-software co-design for regional coalescing of updates and tree-like merging of proxy data using the task-based programming model.
This is efficiently executed utilizing two key hardware components:
\emph{Task-integrated P-caches}, and the \emph{Cascading Router Logic}.
P-caches, in addition to allowing coalescing of updates also contain the logic needed to trigger reduction tasks based on the write-propagation policy of the P-cache.
This efficiently limits the footprint of the data copies when using data-private reductions, while introducing the ability to have asynchronous merging of updates.
The Cascading Router enables \emph{selective cascading}, where invocations sent to the owner tile can be captured and processed at proxies en route based on whether the router port ahead is full or the proxy tile is eager to process it (\cref{sec:cascading}).

Thanks to these innovations, \proj achieves orders of magnitude greater scaling capability for processing sparse applications than previously demonstrated~\cite{dalorex,tesseract}; we evaluate with up to a million PUs across a thousand chips, with no dataset preprocessing or partitioning.

\noindent
\textbf{The technical contributions of this paper are}:
\begin{itemize}
\itemsep0em
\item Hardware-software co-design of a reduction tree approach for task-based parallelization schemes. 
\item Software-configurable proxy region sizes for coalescing and filtering reduction operations (the leaves of the tree).
\item Opportunistic and asynchronous propagation of updates through the tree thanks to the router's selective cascading and P-cache's write-propagation policy.
\item Efficient handling of temporal storage at each level of the tree with P-caches that are integrated into the task-invocation mechanism.
\end{itemize}

\noindent
\textbf{We evaluate \proj and demonstrate that it:}
\begin{itemize}
\itemsep0em 
\item Obtains additive improvements from coalescing/filtering at the proxy regions, and opportunistic asynchronous cascading.
These improvements benefit monolithic mesh and torus networks as well as multichip systems. 
\item Grows performance gains with the parallelization, e.g., 6$\times$ over Dalorex for 16K PUs and 14$\times$ for 64K PUs.
\item Scales well up to 1 million PUs for graphs of a billion edges, while prior work plateaus beyond 10K PUs.
\item Yields 8.6$\times$ higher BFS throughput than the Graph500's top entry for RMAT-26, and 25$\times$ for RMAT-22.
\end{itemize}
\section{Background and Motivation}\label{sec:background}

Memory accesses in graph and sparse linear algebra applications do not exhibit spatial or temporal locality, resulting in poor cache behavior and intense traffic in the memory hierarchy~\cite{graphattack}.
Prior work aiming to accelerate these workloads mitigate memory latency via decoupling, prefetching, and hardware pipelining techniques~\cite{prodigy,maple,graphicionado,graphpulse,ozdal,chronos, pipette, fifer, polygraph, jeffrey_hive, jeffrey_swarm}.
Fifer~\cite{fifer} and Polygraph~\cite{polygraph} increase utilization further through spatio-temporal parallelization.
However, these works are limited in their scalability as they hit the network and memory bandwidth limits already at hundreds of cores.
Prior work aiming to optimize bandwidth includes coalescing~\cite{phi, coup, rich} of updates at various cache levels through dedicated hardware units.
Tesseract~\cite{tesseract} and GraphQ~\cite{graphq} aim to alleviate the memory-bandwidth problem via processing-in-memory, however their proposed integration of PUs on the logic layer of a memory cube~\cite{hmc}, constrains parallelization degree of a given dataset size.
These works demonstrate scaling up to hundreds of PUs.
To scale even further, Dalorex~\cite{dalorex} proposes a task-based parallel execution model on top of tile-based distributed-memory manycores so that each tile holds an equal-sized chunk of the dataset which only the tile's PU can operate on.
This approach makes all data accesses local, allowing Dalorex to scale to thousands of tiles.
However, for larger grid sizes, communication becomes a bottleneck as the longer average distance of task invocations results in higher contention in the network.
In addition, strong scaling results in each tile being responsible for a smaller fraction of the dataset, leading to a statistically more imbalanced work distribution.

\proj offers an efficient software configurable coalescing and reduction-tree approach specialized for distributed-memory manycore systems to minimize communication at large scales by cascading updates.

We detail the data-local execution model of Dalorex in \cref{sec:background_dalorex} which this paper utilizes,
and other relevant software-based approaches in \cref{sec:background_prior}.

\subsection{The Data-Local Execution Model}\label{sec:background_dalorex}

Dalorex co-designs its data-local execution with an architecture that consists of homogeneous tiles, each containing a PU similar to in-order RISC-V cores~\cite{celerity,snitch}, a scratchpad memory, a task scheduling unit (TSU), and a router.
The scratchpad stores a chunk of the dataset and the per-task code and queues.
The network communication is composed of task-invocation messages routed in task-specific channels that share the Network-on-Chip (NoC).
The router has bi-directional ports to North, South, East, West, and the TSU.

\textbf{\emph{Task-based program execution:}}
In Dalorex, the original program is split into tasks that execute at the tile co-located with the memory region (data) that the task operates on.
The resulting program does not have a main execution thread but it is a sequence of tasks that invoke other tasks upon pointer indirection.
A task can spawn other tasks by pushing the inputs for each new task into the PU's output queue (OQ), which the TSU eventually drains into the NoC.
Since the mapping of the dataset arrays to tiles is static after compilation and the first parameter of a task message is a global index to a data array (due to splitting at pointer indirection), this index is used to route the message to the destination tile (the owner of the data)---avoiding headers altogether.
Every tile has an input queue (IQ) per task type, where the TSU places incoming tasks and eventually schedules them for execution at the PU.

\textbf{\emph{Distant task communication:}}
Due to pointer indirection, spawned tasks may target any random tile in the grid, i.e., whoever owns the data to be processed next.
As the size of the grid increases, so would the average number of router hops of a task message and, thus, the network contention.
\proj introduces \emph{proxy ownership} and \emph{selective cascading} in \cref{sec:approach} to coalesce and merge data through asynchronous reduction trees, thereby decreasing distant communication in large tile grids.

\subsection{Relevant Software Approaches}\label{sec:background_prior}

While it is possible to implement a reduction tree approach managed in software, this presents significant limitations such as synchronization~\cite{mpich,sw_reduce} or otherwise active waiting to handle messages, which are not managed efficiently in general-purpose software systems~\cite{openmp,mpi}.
In addition, storage requirements for software-managed reductions are often the size of the entire array to be reduced, times the number of copies.
\proj offers asynchronous and opportunistic merging of reductions through the P-cache and decreases the storage overhead by orders of magnitude over software-based approaches (\cref{sec:proxy_cache}).

Another related software-based approach is Gluon~\cite{gluon,gluon_async} which offers a lightweight API to enable optimization of communication when running programs on distributed systems that process partitioned graph data.\footnote{Data partitioning is a preprocessing step used in distributed graph processing to minimize cross-node communication~\cite{giraph++,metis}. Our approach is orthogonal to data partitioning as it could also be applied within each partition.}
Gluon's approach is complementary to \proj and we expect an additive effect if they were to be combined.

\definecolor{codegreen}{rgb}{0,0.6,0}
\definecolor{codegray}{rgb}{0.5,0.5,0.5}
\definecolor{codepurple}{rgb}{0.58,0,0.82}
\definecolor{backcolour}{rgb}{0.97,0.97,0.96}
\lstdefinestyle{mystyle2}{
    backgroundcolor=\color{backcolour},   
    commentstyle=\color{codegreen},
    keywordstyle=\color{blue},
    numberstyle=\tiny\color{codegray},
    stringstyle=\color{codepurple},
    basicstyle=\ttfamily\scriptsize,
    breakatwhitespace=false,         
    breaklines=true,                 
    captionpos=b,                    
    keepspaces=true,                 
    numbers=left,                    
    numbersep=3pt,                  
    showspaces=false,                
    showstringspaces=false,
    showtabs=false,                  
    tabsize=2,
    morekeywords={}
}

\section{The Tascade Approach}\label{sec:approach}

As aforementioned, \proj aims to minimize the communication distance of reduction tasks and the load imbalance across the PUs and the NoC that arises from large-scale parallelization of sparse applications.
Since these applications perform associative and commutative reduction operations, \proj puts forward hardware support for efficient and scalable reductions.
Building on a similar task-based execution model as Dalorex, \proj relaxes the single-owner-per-data constraint for reduction arrays that results in two modes of ownership.

\textit{\textbf{Data ownership.}}
As in Dalorex, the dataset arrays are distributed across the address space so that every tile \emph{owns} an equal-sized chunk of each data array.
In the context of reduction tasks, we use \emph{data owner} to refer to the tile that owns the right to load-store a particular element of the reduction array ($R_{array}$) which is stored physically in the same tile. 

\begin{figure}[t]
\centering  
\includegraphics[width=\columnwidth]{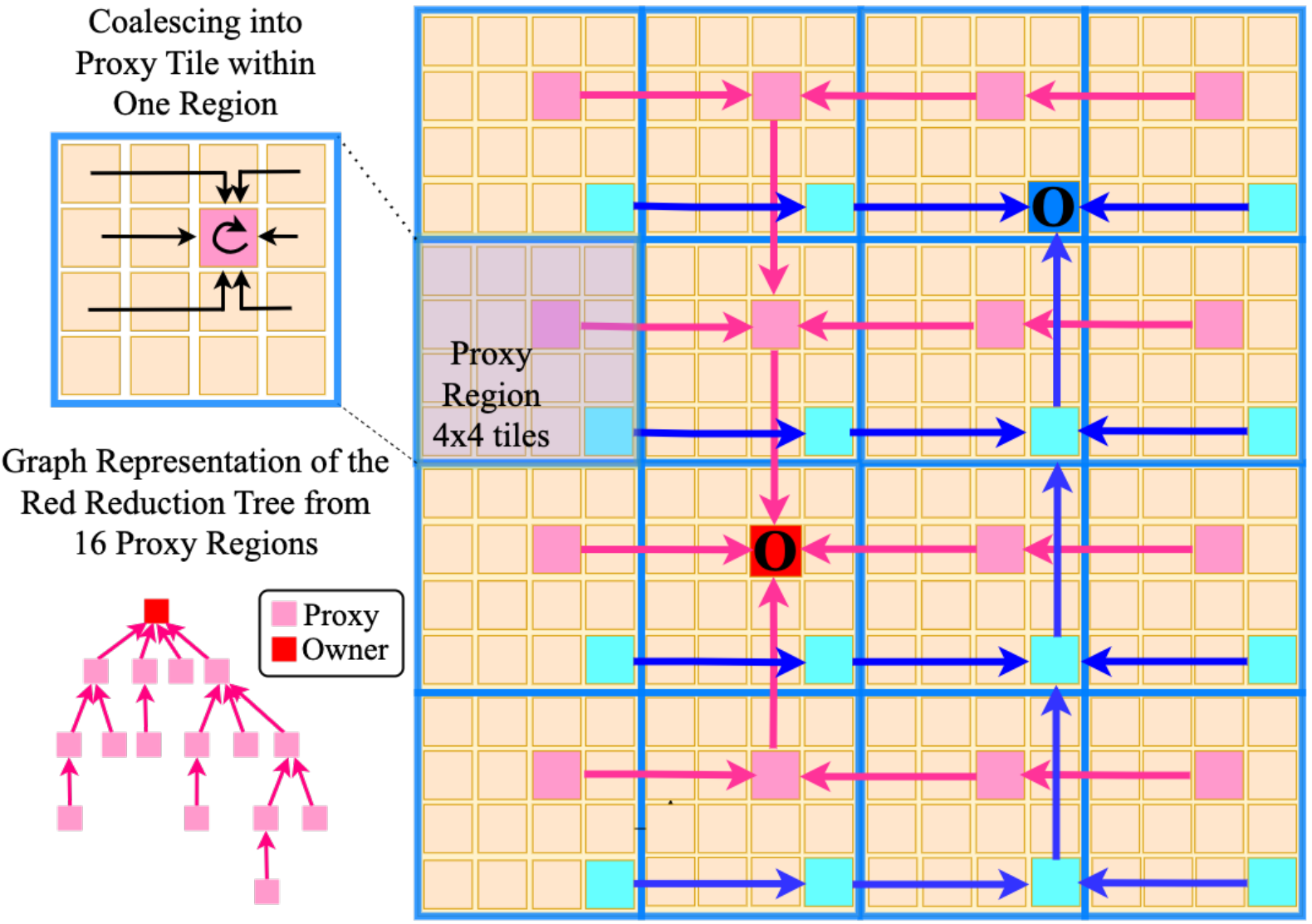}
\vspace{-1mm}
\caption{Reduction Trees of \proj.
\textbf{Right:} We depict the flow of updates for a 2D mesh topology for two different reduction trees, where the blue and red tiles are the roots (data owners), and the cyan and magenta tiles are the corresponding proxy tiles.
Since all proxies of a particular element of the reduction array have the same coordinates within each region, this allows for cascading updates selectively by the proxies en route to the owner (O).
\textbf{Left:} Flow of updates for coalescing within a proxy region prior to reduction (top). Graph representation of one of the reduction trees (bottom).}
\vspace{-5mm}
\label{fig:reduction_tree}
\end{figure}

\textit{\textbf{Proxy ownership.}}
This additional mode emerges by allowing storage of a temporary copy of the reduction array, called \emph{proxy array} ($P_{array}$), per subdivision of the tile grid, called \emph{proxy region} (see \cref{fig:reduction_tree}).
In a given proxy region, a \emph{proxy tile} is responsible for caching elements belonging to a fraction of the $P_{array}$---equally divided among the region's tiles.
Having each $P_{array}$ distributed across a region---as opposed to each tile having its own---decreases the storage overhead of data-private reductions by a factor of $W^2$ where $W$ is the width of the proxy region---assuming square-shaped regions.
We say that a tile is \emph{a proxy for} the elements of the $R_{array}$ for which it owns the $P_{array}$ counterpart, as it can operate on those elements, and eventually merge the updates to the data owner.
A tile operates on the $P_{array}$ via the Proxy cache (P-cache) to further decrease the storage overhead of data-private reductions by $C$, i.e., the ratio of the local $P_{array}$ fraction to the P-cache size (detailed in \cref{sec:proxy_cache} and evaluated in \cref{fig:proxy_pressure}).

In \proj, each tile can be viewed as the root of a reduction tree for the $R_{array}$ elements it owns.
The rest of the nodes in the tree are the proxies of those elements, one per proxy region.
\cref{fig:reduction_tree} depicts that for two separate elements for which the blue and red tiles are the owners, and the cyan and magenta tiles are its proxies.
The tree is asymmetric since the proxies are distributed across the grid, one per region, on the same coordinates within each region as the data owner.
The rationale for this is that since 2D NoC topologies often use dimension-ordered routing, the proxy tiles are on the path of the data updates toward the owners and so they can capture these updates as proxy tasks (see \cref{sec:proxy_tasks}) to filter and coalesce them, minimizing the NoC traffic.
In addition, to improve PU and NoC utilization, the updates are captured opportunistically, leveraging our selective cascading approach (\cref{sec:cascading}).

\textbf{\textit{Applicability:}}
\proj can be applied to manycores architectures regardless of whether they have a fully distributed memory across the tiles (e.g., Dalorex~\cite{dalorex} or Cerebras~\cite{cerebras_arch,cerebras_fft}, where each tile only accesses its local memory) or use a partitioned global address space (PGAS) where tiles can access remote memory~\cite{celerity}.
For generality, throughout the rest of the paper, we view the global address space as a concatenation of the local address spaces that each tile is responsible for.
The local address space does not need to fit the tile's local SRAM memory size, as this memory could be used to cache a portion of a larger address range mapped to an external DRAM memory~\cite{dcra}.
Having that in mind, we consider two modes in which dataset arrays can be accessed, as a data cache (D-cache) or as a scratchpad.
The relative size of the dataset arrays and the tile's memory can be considered when deciding the size of the P-cache, with a heuristic we elaborate on \cref{sec:proxy_cache_config}.

\subsection{Proxy Tasks}\label{sec:proxy_tasks}

\begin{figure}[t]
\centering  
\includegraphics[width=\columnwidth]{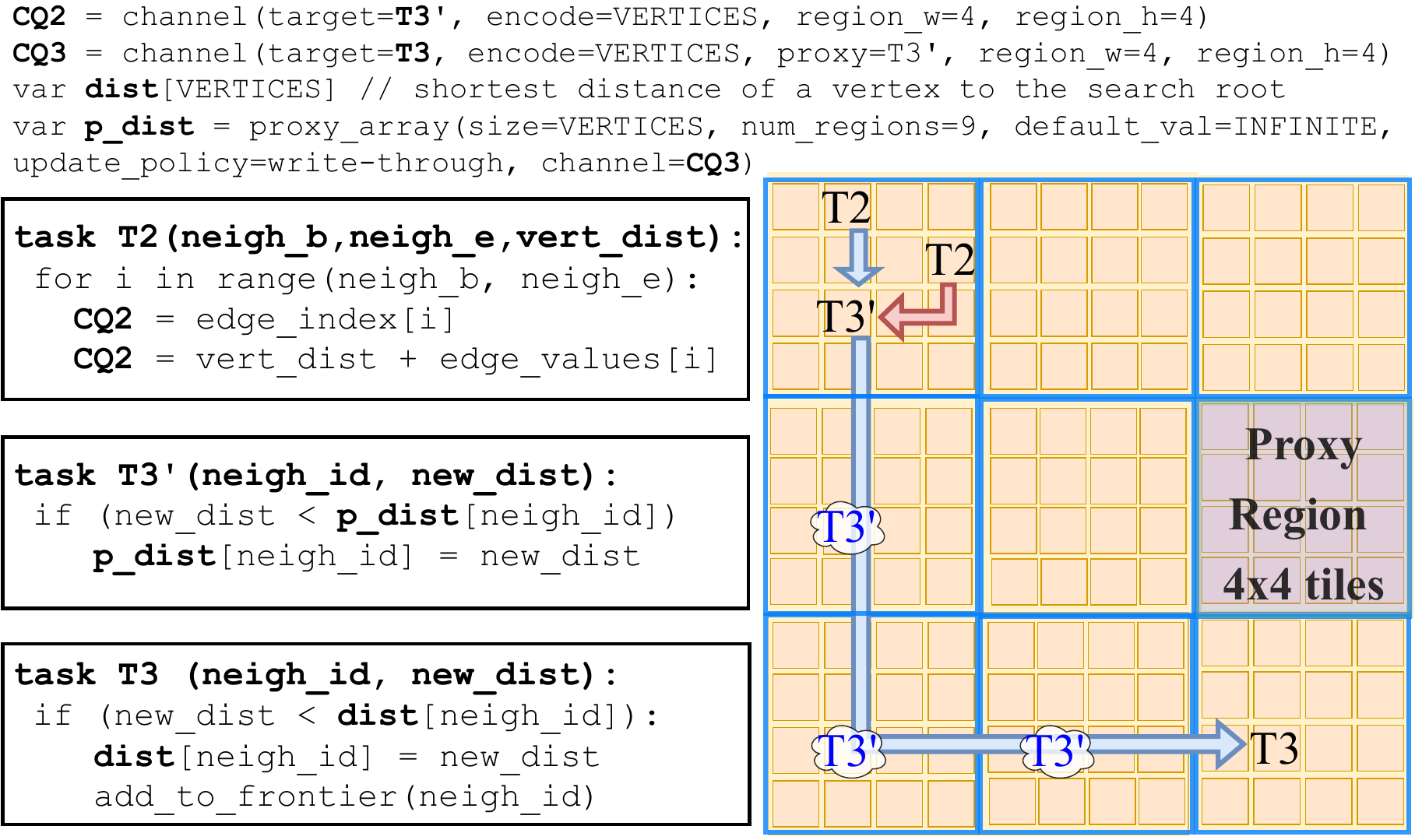}
\vspace{-6mm}
\caption{
Tasks for single-source shortest path (SSSP) and configuration of proxy regions (of 4x4 tiles) in \proj.
Each region operates on its copy (\texttt{p\_dist}) of the reduction array (\texttt{dist}) where each tile \emph{is a proxy for} a fraction of it.
Contrary to Dalorex, in \proj, \texttt{T2} invokes \texttt{T3'}, which targets a proxy tile within the same region.
\texttt{T3'} tasks write to \texttt{p\_dist}, which lives in the P-cache (see \cref{sec:proxy_cache}).
Upon a cache miss, it returns the default value.
\texttt{T3} tasks are invoked based on the write-propagation policy of the P-cache (i.e., write-through in SSSP).
The \textcolor{red}{red arrow} shows a \texttt{T3'} that does not result in an update in P-cache.
The \textcolor{blue}{blue arrow} shows a \texttt{T3'} that updates the P-cache and triggers a \texttt{T3} invocation towards the owner tile.
Through selective cascading, en route to the owner, this \texttt{T3'} message may be captured by a proxy tile---indicated with clouds.
}
\vspace{-5mm}
\label{fig:proxy_init}
\end{figure}

Any reduction task in the data-local execution model can have a proxy task (i.e., a task that operates on a copy of the $R_{array}$).
Our evaluation applies this to the vertex update task of graph applications and the output vector for histogram and SPMV (\cref{sec:results})
\cref{fig:proxy_init} exemplifies this on the single-source shortest path (SSSP) algorithm
where \texttt{T3} is the reduction task and \texttt{T3'} is its proxy task.

The \textbf{\textit{configuration code}} in \cref{fig:proxy_init} (above) describes how proxy regions are defined in software with a grid example of 9 proxy regions of 4x4 tiles.
The proxy task (\texttt{T3'}) operates on the proxy array (\texttt{\texttt{p\_dist}}), that is, each region's copy of the reduction array (\texttt{\texttt{dist}}).
Reads and writes to the proxy array are directed to the P-cache, and so defining \texttt{\texttt{p\_dist}} also configures the P-cache.
This configuration includes the default value returned upon cache misses and the policy for propagating updates (elaborated on \cref{sec:proxy_cache}), as well as the channel that will carry that update.

The write-propagation policy determines how the updates made to the P-cache are propagated to the tile that owns the corresponding elements of \texttt{\texttt{dist}}.
The SSSP code in \cref{fig:proxy_init} uses a write-through policy, where updates are immediately propagated to the owner tile.
However, because SSSP performs a minimization operation \texttt{T3'} tasks only write to the P-cache if the \texttt{new\_dist} value is smaller than the current value.

\cref{fig:proxy_init} depicts a \textbf{\textit{filtering scenario}} with two invocations of a proxy task (\texttt{T3'}) executed on the same tile but coming from different \texttt{T2} tasks, where only one of them updates the P-cache.
First, this scenario considers a \texttt{T2} task invoking (\textcolor{red}{red arrow}) a \texttt{T3'} for which \texttt{new\_dist} is larger than the \texttt{p\_dist[neigh\_id]} value stored in the P-cache, and thus it is not updated; we refer to this as filtering since having a local proxy avoided the long-distance communication to the owner tile.
Second, another \texttt{T2} invokes (\textcolor{blue}{blue arrow}) a \texttt{T3'} that does update \texttt{p\_dist} and thus, the P-cache propagates the update using the configured channel (\texttt{CQ3}).
Next, this update message---addressed to execute \texttt{T3} at the owner (as per \texttt{CQ3} configuration)---may be captured by one of the proxy tiles while en route to the owner tile, based on the selective cascading policy (\cref{sec:cascading}).
If captured, then \texttt{T3'} is executed on the proxy tile, which may or may not cause an update, that would then continue its way to the owner tile.

The reader may have realized that in this minimization-based reduction, the regions closer to the owner are more likely to have more up-to-date values in their P-caches that will filter out updates from further regions.
This is different for addition-based reductions (e.g., Histogram) where the P-caches coalesce regional updates that are eventually propagated to the owner tile on eviction.
In either case, data updates are propagated asynchronously and transparently to the software, thanks to the P-cache design, which we detail in the next section.

\vspace{-0.5mm}
\subsection{Proxy Cache Design}\label{sec:proxy_cache}
\vspace{-0.5mm}

Since the P-cache is configured in software when defining a $P_{array}$, we make a design that does not introduce significant area overhead when the P-cache is not configured.
To do that, a portion of the tile's local SRAM memory is utilized as a direct-mapped cache (associativity=1), to which the accesses to the local $P_{array}$ fraction ($P_{array\_local}$) are directed.
The P-cache stores cacheline tags and the valid bit in SRAM too, so the area overhead of the cache is only the logic gates for tag comparison and configuration registers (see \cref{sec:proxy_cache_config}).
This design is aligned with trends in manycore architectures where the per-tile memories are used as scratchpads or as configurable data caches~\cite{cerebras_arch,dcra,decades}.
Note that since the footprint of the P-cache ($P_{array\_local}$) is known at compile time, the number of bits needed for the tag is $\log_2$ of the ratio between the $P_{array\_local}$ elements and the SRAM available to them (\cref{sec:proxy_cache_config} elaborates on how this SRAM portion is determined).

\textbf{\textit{P-cache misses and evictions:}}
A miss in the P-cache returns a preconfigured default value.
This would correspond to the initial value of the reduction array, e.g., zero for addition or maximization, or infinite for minimization.
On eviction, the data is either sent as a task invocation to the owner tile (write-back mode) or ignored (write-through).
This reduces NoC traffic over software-based reductions where the copies of the $R_{array}$ are stored in the data cache, and cache misses and evictions must traverse the memory hierarchy.

\textbf{\textit{Write-propagation policy:}}
Upon a cacheline being updated or evicted, the data address and value are sent as a message to the data owner.
To enable this, the P-cache---similar to the PU---has the ability to push into a network channel via the output queues (OQs).
For simplicity, a cacheline contains a single data element to avoid sending multiple messages.
To avoid additional buffers, the TSU ensures that the OQ has sufficient space for the P-cache to push a task invocation before scheduling any task that may result in P-cache eviction or update.

\textbf{\textit{Write-back}} enables coalescing updates to the same cacheline and only sending the aggregated data to the owner tile upon eviction.
The P-cache also self-invalidates cachelines when the PU is idle and all its OQs are empty.
This policy enables the merging of the proxy values to the owner tile (resembling the reduction tree), asynchronously and opportunistically, throughout the program execution, without having to wait for the end of the computation phase to merge the results.
This mode \textbf{\textit{enables update coalescing}}, which is suited for time-insensitive reductions that have either a single computation phase (e.g., Histogram) or a barrier between search epochs (e.g., Pagerank).

\textbf{\textit{Write-through}}, alternatively, has every data update triggering a task invocation towards the owner tile.
This mode \textbf{\textit{enables data filtering for minimization or maximization operators}}, as only a new minimum or maximum value writes to the P-cache.
Write-through allows the updates to reach the owner tile as soon as possible to minimize the redundant explorations of vertices in the frontier.
This mode is suitable for barrierless implementations of graph applications~\cite{barrierless}, like the SSSP code in \cref{fig:proxy_init}. 
With write-through, there is no need for the P-cache to self-invalidate, as the updates are always pushed to OQs to make their way to the owner tile.

\vspace{-0.5mm}
\subsection{Proxy Cache Configurations}\label{sec:proxy_cache_config}
\vspace{-0.5mm}

The P-cache has five memory-mapped configuration registers.
The compiler would have heuristics to set these registers, which can be overridden by the programmer if desired.
\begin{enumerate}
\item \textbf{Local proxy array fraction}: set to $P_{array}/W^2$ where $W$ is the width of a proxy region---determined by \cref{eq:W_min}.
It determines the range of the tile's local address space for which memory operations are directed to the P-cache.
\item \textbf{P-cache size}: set by \cref{eq:P_cache}. It determines the chunk of the tile's SRAM that is reserved for the P-cache.
\item \textbf{Write-propagation policy}: set based on the application needs for data timeliness, as described above.
\item \textbf{Channel ID to propagate updates}: set to the channel that routes to the reduction task (e.g., \texttt{T3} in \cref{fig:proxy_init}).
\item \textbf{Default value} for cache misses: set based on the type of reduction operation, which could potentially be detected by the compiler.
\end{enumerate}

To ease the programmer's burden for \textit{\textbf{determining the proxy region and P-cache sizes}}, we created the following heuristic, which would be set by the compiler but can be overridden by the programmer if desired.
Since $P_{array\_local}$ (i.e., the footprint of the $P_{array}$ on the P-cache) is $P_{array}/W^2$, where $W$ is the width of a proxy region (assuming square regions), using $C$ as the maximum desired ratio between $P_{array\_local}$ and the maximum SRAM size to be dedicated to the P-cache ($P_{cache\_max}$), we compute the smallest region that can be configured $W_{min}$ as:
\begin{equation}\label{eq:W_min}
    W_{min} = \sqrt{\frac{P_{array}}{P_{cache\_max} \cdot C}}
\end{equation}

We studied various values of $C$ in \cref{fig:proxy_pressure} and found $C=16$ to maintain most of the performance benefits of the P-cache.
The value for $P_{cache\_max}$ depends on the system integration and the ratio of the dataset arrays to the tile's SRAM.
For example, on a system where all its memory is distributed across the tiles (e.g., Dalorex or Cerebras), $P_{cache\_max}$ would be set to use all the free SRAM on the tile.
That could also be the policy for systems with an external memory backing up the local memories but when the parallelization is such that the dataset arrays fit on the SRAM.
On the contrary, when the local memory is used as a D-cache, $P_{cache\_max}$ would be set to a modest fraction of the SRAM.

Note that \cref{eq:W_min} outputs the smallest region size that can be configured, for which $P_{array\_local}$ would be the largest ($C$ times larger than the maximum P-cache size configurable).
However, the selected proxy region size could be larger, e.g., $W$=16, which was experimentally found to be a good balance between the P-cache size and the proxy region size (\cref{fig:proxy_comp}).

Thus, we could calculate the actual P-cache size as:
\begin{equation}\label{eq:P_cache}
    P_{cache} = \min(\frac{P_{array}}{(\max(16, W_{min}))^2}, P_{cache\_max})
\end{equation}

\subsection{Cascading}\label{sec:cascading}

Since proxy arrays are distributed across each region in the same way, the proxy tiles for a particular $R_array$ element are on the same row/column for horizontally/vertically aligned proxy regions.
Therefore, when a task invocation moves towards the owner tile across the NoC in a dimension-ordered manner, it will naturally pass by its corresponding proxy tiles en route.
This is depicted in \cref{fig:reduction_tree} for two different owner tiles on a 2D mesh.

As a task invocation passes by a proxy tile, the router can capture it as a proxy task and execute it on the local PU.
Based on the task execution on write-through (i.e., whether the new value is a minimum or maximum) or upon a P-cache eviction on write-back, a new task is spawned, which may be captured by the next proxy tile en route; we call this process \emph{cascading}.

There are two modes for en route proxy task processing:
(1)~\emph{Always Cascading}: Every proxy tile en route (i.e., one per proxy region) is obliged to process the task.
(2)~\emph{Selective Cascading}: A proxy tile opportunistically decides whether to process the task based on the occupancy of its IQ or the contention on the router's output port ahead.
If the IQ occupancy is less than half of its capacity, or the network ahead is congested, the tile captures the task. Otherwise, it lets the task continue towards the owner tile.

\cref{lst:selective_cascading_code} shows the logic that we added to the router to support cascading and its selective mode.
We added two configuration registers (for coordinates X and Y) to store the masks of the bit selection that determines whether the tile is a proxy owner for a given message (line 7) and a register to enable/disable proxy usage.
For every incoming message, the router determines if the current tile is the owner or proxy tile of the data (lines 14-21).
If it is neither, the router moves the data in the direction of the owner tile.
If it is the owner tile, then it directs the data to the corresponding task's IQ (\texttt{T3} in \cref{fig:proxy_init}).
Alternatively, if it is a proxy tile, it may capture the message into the proxy task's IQ (\texttt{T3'} in \cref{fig:proxy_init}), based on the occupancy of the IQ and the buffer of the outgoing network port (lines 9-12).

\begin{lstlisting}[escapeinside={(*}{*)}, language=Verilog, caption={
Snippet of the Verilog code needed to implement proxy regions and selective cascading.
The proxy region is configured by setting the proxy\_mask and proxy\_enable registers (i.e., flip flops).
The id\_x\_within and id\_y\_within registers are the coordinates of the tile within the proxy region.
The select\_msg wire determines whether the message should be captured by the proxy region or let through.
The sequential depth of select\_msg is not worse than the proxy comparison, which is on par with the sequential depth of is\_dest.
Therefore, the critical path of route\_to\_core is only an OR gate more than the existing logic.
Note that the critical path of the $>=$ operators involved in calculating the cardinal directions is longer than the is\_dest logic, and thus, our addition is probably not affecting the overall critical path in most router designs.
}, label=lst:selective_cascading_code, style=mystyle2]
input [15:0] dest_x, dest_y;
input [1:0] input_port; // N,S,E,W
reg [15:0] id_x, id_y; // Existing Tile ID Registers
// Proxy Configuration Registers
reg proxy_enabled_r; // To enable usage of proxy regions
reg [3:0] proxy_mask_x, proxy_mask_y; // 4'b0011 for 16x16

// Whether the opposite-facing port from the inputs N,S,E,W had its buffer full last cycle, and proxy is enabled
reg [1:0] opposite_port_buffer_full_r;
// The occupancy of the input queue of the PU is <= half
reg PU_IQ_lt_half_full_r; // It includes proxy_enabled_r
wire select_msg = PU_IQ_lt_half_full_r || opposite_port_buffer_full_r[input_port];
//We flop id_within to remove a gate from the critical path
reg [5:0] id_x_within ={id_x[5:2] & proxy_mask_x,id_x[1:0]}
reg [5:0] id_y_within ={id_y[5:2] & proxy_mask_y,id_y[1:0]}
wire is_proxy_x = {dest_x[5:2] & proxy_mask_x, dest_x[1:0]} == id_x_within;
wire is_proxy_y = {dest_y[5:2] & proxy_mask_y, dest_y[1:0]} == id_y_within;
// Sequential depth: 6-bit comparator + three AND
wire go_to_proxy = is_proxy_x && is_proxy_y && select_msg;
// Destination logic: 16-bit comparator + AND
wire is_dest = (dest_x == id_x) && (dest_y == id_y);
// Adding an OR gate to the critical path of is_dest
wire route_to_core = go_to_proxy || is_dest;
\end{lstlisting}

\vspace{+2mm}
\textit{\textbf{Lightweight router additions:}}
As described in \cref{lst:selective_cascading_code} the logic for identifying a tile as a proxy for a message is done in parallel with the logic that determines whether the tile is the destination, and thus, we only add one OR-gate to this path (line 25).
The critical path of calculating \texttt{go\_to\_proxy} is not longer than the destination \texttt{is\_dest} logic since the proxy-mask comparator employs fewer bits, and \texttt{select\_msg} is determined without using the message destination.
Counting the logic added to the router, \cref{lst:selective_cascading_code} shows an addition of 20 flip-flops and a few dozen logic gates.
This represents a negligible overhead to the overall area of the router given its complex per-port multiplexing logic and message buffers.

\section{Evaluation Methodology}\label{sec:methodology}

We use MuchiSim~\cite{muchisim} to evaluate \proj as it was the simulation infrastructure used in Dalorex, against which we compare.
MuchiSim is a functional simulator that models the NoC in a cycle-accurate manner---a level of detail that is crucial to evaluate \proj innovations.
We extended the simulator to model proxy caches and cascading, and validated its functional correctness by comparing the application results.

We consider two different \textbf{\textit{system integrations}} for our experiments:
(1) a large monolithic grid of tiles that fits the entire dataset on-chip, as in the Dalorex paper~\cite{dalorex} or the Cerebras Wafer Scale Engine~\cite{cerebras,cerebras_hotchips}; and
(2) a multi-chip system where every 32x32-tile chip is attached to a 12 GiB HBM2E DRAM with eight 64 GB/s memory channels.
The latter is employed for the multi-chip experiments of \cref{fig:sync_mesh} and \ref{fig:scaling_million}, while the former is used for all other experiments to compare with Dalorex on their proposed architecture.

\textbf{\textit{Parameters:}}
When comparing with Dalorex, we use the same configuration for \proj so that the only hardware difference is our additions to support the P-cache and cascading.
In all the experiments, we configure a 1 Ghz frequency at 0.7 V with 7 nm technology and select the amount of SRAM per tile to 512 KiB and the network-on-chip (NoC) width to 64-bit.
The NoC is a 2D torus (as in Dalorex) in all experiments except for the one characterizing network topologies that also evaluates a 2D mesh (\cref{fig:sync_mesh}).
In the multi-chip case, we evaluate a 2D torus topology across chips and use the default parameters of MuchiSim for inter-chip latency and energy (20 ns~\cite{pcie6} and 1.17 pJ/bit~\cite{nvidia_chiplets}).
Full details of the latency and energy parameters used as well as all the simulation logs can be found in our \repository.

\textbf{\textit{Applications:}}
In addition to the four graph algorithms evaluated by Dalorex~\cite{dalorex}---Breadth-First Search (BFS), Single-Source Shortest Path (SSSP), PageRank (PAGE), and Weakly Connected Components (WCC)---we also evaluate Sparse Matrix-Vector Multiplication (SPMV) and Histogram, to demonstrate the generality of our approach for data-intensive applications.

\textbf{\textit{Datasets:}}
We use three sizes of the RMAT graphs~\cite{kron}---standard on the Graph500 list~\cite{graph500}---RMAT-22, RMAT-25 and RMAT-26, which are named after their number of vertices, e.g. RMAT-26 (R26) contains $2^{26}$ (67M) vertices (V) and 1.3B edges (E) with a memory footprint of 10.6 GiB.
We also use the Wikipedia (WK) graph (V=4.2M, E=101M) in our evaluation to evaluate different graph topologies.
For SPMV we use the same datasets, since a graph is a square sparse matrix of $V\times{V}$ with $E$ non-zero elements.
The graph data is stored in Compressed Sparse Row (CSR) format \cite{csr} without any partitioning, resulting in three input arrays. 
The output array has size $V$ in each case.

\subsection{Comparing with the State-of-the-Art}\label{sec:graph500}

We use Dalorex as the SotA system for comparison because we target a similar task-based parallelization scheme and because it has shown scalability for graph processing on a publicly available detailed architectural design.

In addition, we evaluate the performance of \proj compared to the Graph500 entries~\cite{graph500}, adhering to their guidelines to the best of our ability, for RMAT-22 and RMAT-26.
Graph500 requires timing separately the reading, preparing, and loading of the graph onto the system from the graph traversal itself.
In our case, we do not perform any dataset pre-processing and directly read the CSR structure from disk.
Based on the Graph500 guidelines, we begin measuring runtime when the search key is loaded onto the system, and we stop when the last vertex is visited.
We report traversed edges per second (TEPS) as $E/time$ where $E$ is the number of edges connected to the vertices in the graph traversal starting from the search key.\footnote{We report TEPS for one search key (id=0), as opposed to random sampling and averaging time across 64 search keys, due to long simulation time.}
Since we evaluate other workloads than graph traversal, we consider the non-zero elements of the sparse matrix for SPMV, and the input array for Histogram as $E$ when reporting TEPS.

\subsection{Evaluating The Impact of Design Contributions}

In addition to evaluating throughput and energy efficiency, we characterize the performance impact of the key hardware innovations introduced in this paper:
(1)~\emph{proxy regions}, with a range of region and P-cache sizes, over the baseline of no proxy;
(2)~\emph{selective cascading} strategy, compared to always or never cascading; and
(3)~\emph{asynchronous merging of reductions} over having a global synchronization prior to merging the proxy (copies of) data.

We also evaluate the aggregated benefits of the \proj innovations for three \emph{network interconnect options}, i.e., 2D-torus and 2D-mesh in the monolithic experiments, and a hierarchical torus for the multi-chip experiments.
These evaluations study runtime and energy efficiency improvements, as well as the impact on the network traffic, which helps to understand the source of the performance gains.

\textit{\textbf{Area overhead:}}
Considering the bits of extra storage as a metric for the overhead of \proj, we obtain 60 bits for the P-cache configuration registers described in \cref{sec:proxy_cache}.
(No additional storage is necessary for the P-cache, as it shares the SRAM pool with the data cache.)
Adding that to the 20 bits counted from \cref{lst:selective_cascading_code} it sums to $\sim 10$ Bytes of registers and some few hundred logic gates per tile.
Considering just the area of local memory, the overhead introduced by \proj of 0.001 KiB is negligible.
The extra logic required is also tiny compared to the 0.2 mm$^2$ area per tile reported by MuchiSim.

\section{Results}\label{sec:results}

\begin{figure}[htp]
\includegraphics[width=\columnwidth,height=2.9cm]{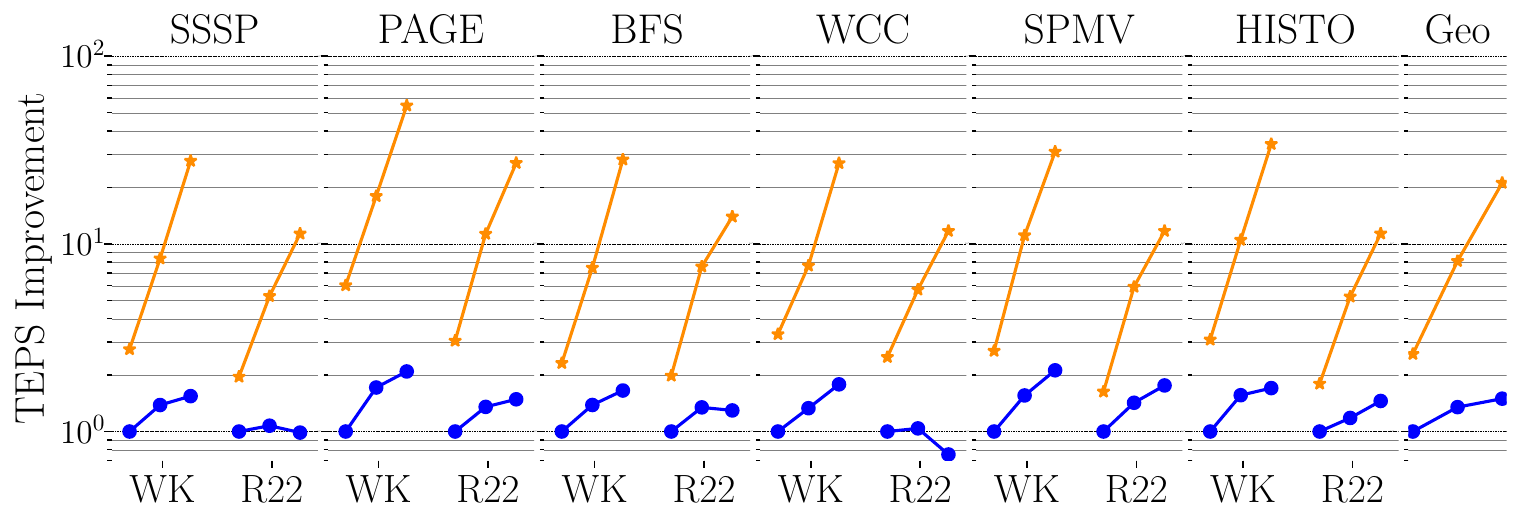}
\includegraphics[width=\columnwidth,height=3.3cm]{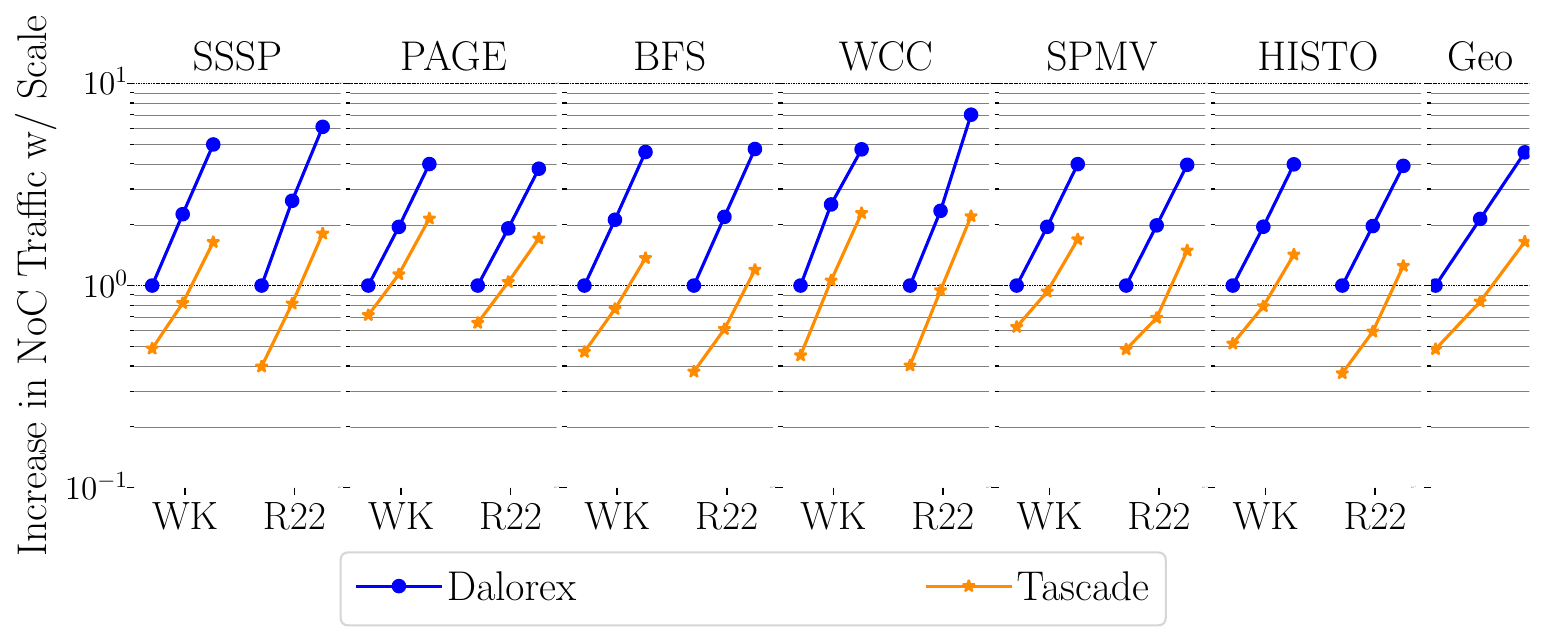}
\vspace{-8mm}
\caption{
Performance gain and network traffic of Dalorex and \proj for three scaling steps for each dataset---64x64 ($2^{12}$ tiles), 128x128 ($2^{14}$) and 256x256 ($2^{16}$) grid---normalized to Dalorex 64x64.
}
\vspace{-3mm}
\label{fig:dalorex_comparison}
\end{figure}

\cref{fig:dalorex_comparison} shows the evaluation of Dalorex and \proj scaling across grid sizes of $2^{12}$ ($64\times64$), $2^{14}$ ($128\times128$) and $2^{16}$ ($256\times256$) tiles processing RMAT-22 and WK datasets. 
The starting point of this scaling experiment is the minimum grid size that can hold the entire dataset on SRAM in this monolithic architecture (with 512 KiB per tile).
The bottom plot showcases the steep increase in NoC traffic due to the longer task-invocation average distance as the parallelization grows.
\textbf{The decreased traffic in \proj allows it to continue scaling performance, while Dalorex performance plateaus demonstrating the impact of the \proj communication-reduction techniques that enable performance scaling.}

\proj demonstrates scalability even with the $256\times256$ grid, i.e., over 65,000 tiles ($2^{16}$) for RMAT-22, which has $2^{22}$ vertices, achieving a massive level of parallelization.
The key features of \proj that enable this are:
(1)~Coalescing and filtering of updates to distant data---via the P-caches---coupled with asynchronous task invocation for sending these updates to the owner tile, and
(2)~Cascading the reduction operations sent to the owner tile through proxy tiles en route which is equivalent to concurrent asynchronous reduction trees across the grid.
The P-cache and the cascading router logic are the key hardware contributions that enable these two features, and we characterize their impact later in this section.

Next, in \cref{sec:res_proxy} we characterize the contribution of P-cache-mediated filtering and coalescing in local proxy regions with no cascading, then show the additional contribution of cascading. 
We then study the impact of the proxy region size in \cref{sec:res_proxy_region_size}, and analyze the sensitivity of performance to P-cache size in \cref{sec:res_proxy_cache_size}. 
\cref{sec:asynchrony} evaluates adding synchronization before merging the proxy updates, measuring the benefit of asynchrony and providing an upper bound for the performance of a software-managed proxy.
\cref{sec:res_noc} evaluates the improvements that \proj provides with different NoC topologies and showcases its applicability to multi-chip systems.
Finally, \cref{sec:million} studies strong scaling (by parallelizing RMAT-26 for grid sizes ranging from a thousand to a million tiles). 

\subsection{Proxy Caching and Cascading Improve Performance}\label{sec:res_proxy}

The single-owner-per-data scheme of Dalorex starts to show sub-linear performance at thousand-tile scales.
In addition to the increased task-invocation distance, the performance degradation is also caused by the work imbalance that grows with the parallelization level.
Since Dalorex requires tasks operating on a given data chunk to be handled by a single tile, the more tiles allocated to process a given dataset, the smaller the chunk of the dataset each tile processes, and the higher the hotness variance across them.

\begin{figure}[!htp]
\includegraphics[width=\columnwidth,height=2.9cm]{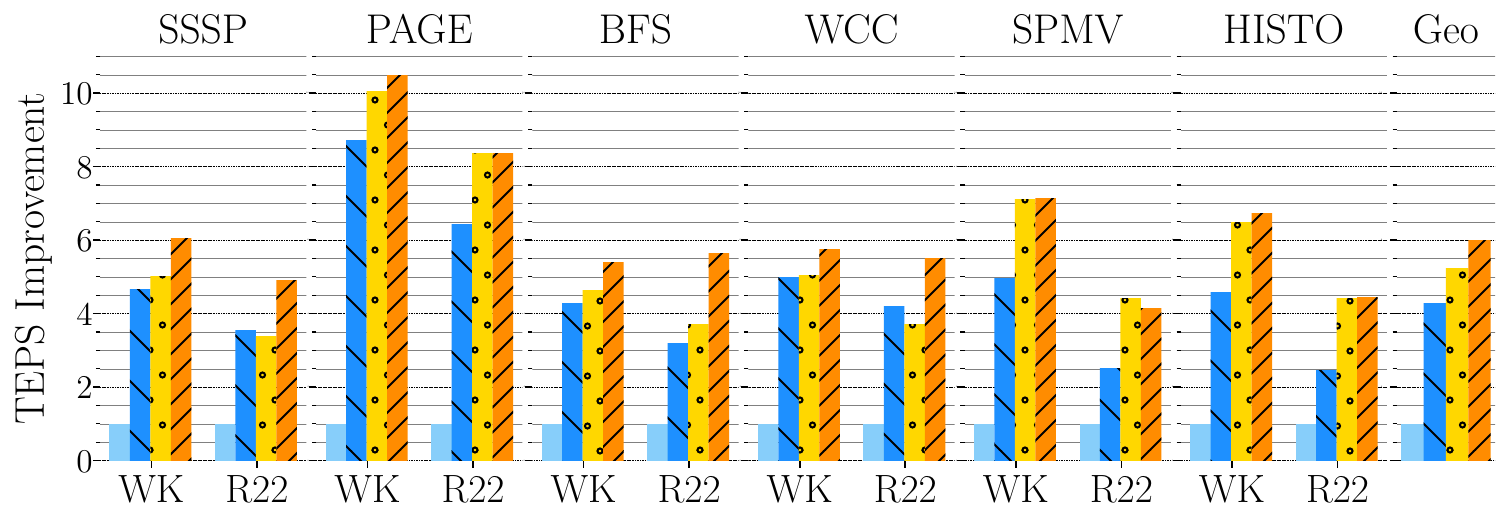}
\includegraphics[width=\columnwidth,height=2.9cm]{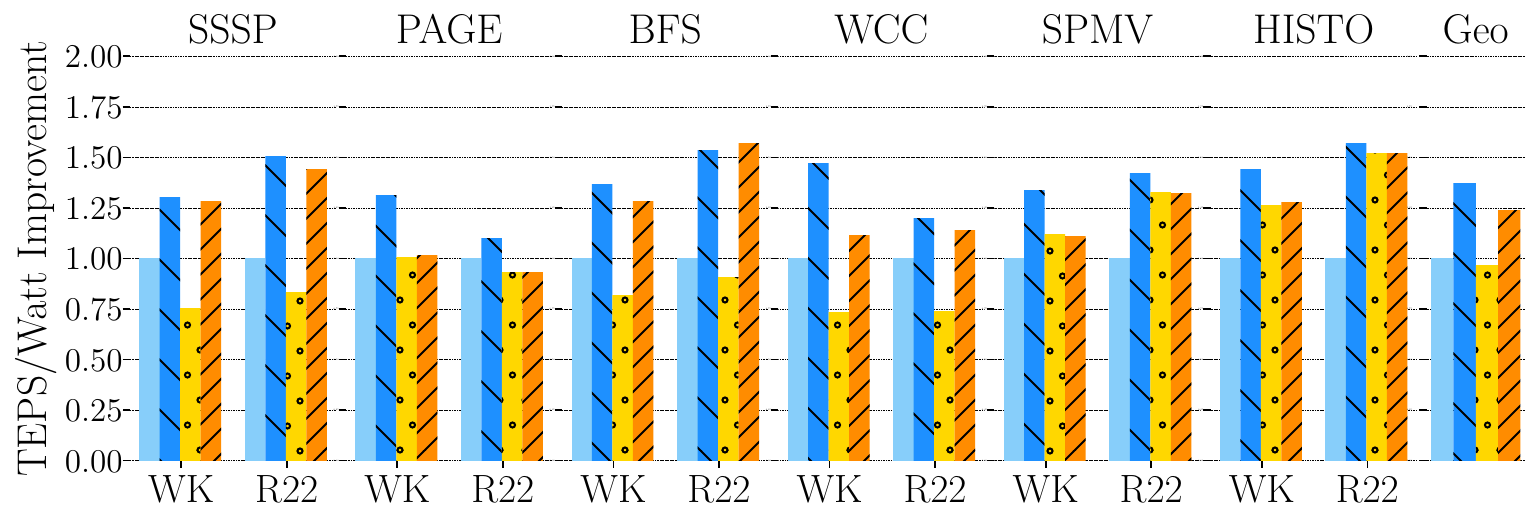}
\includegraphics[width=\columnwidth,height=3.5cm]{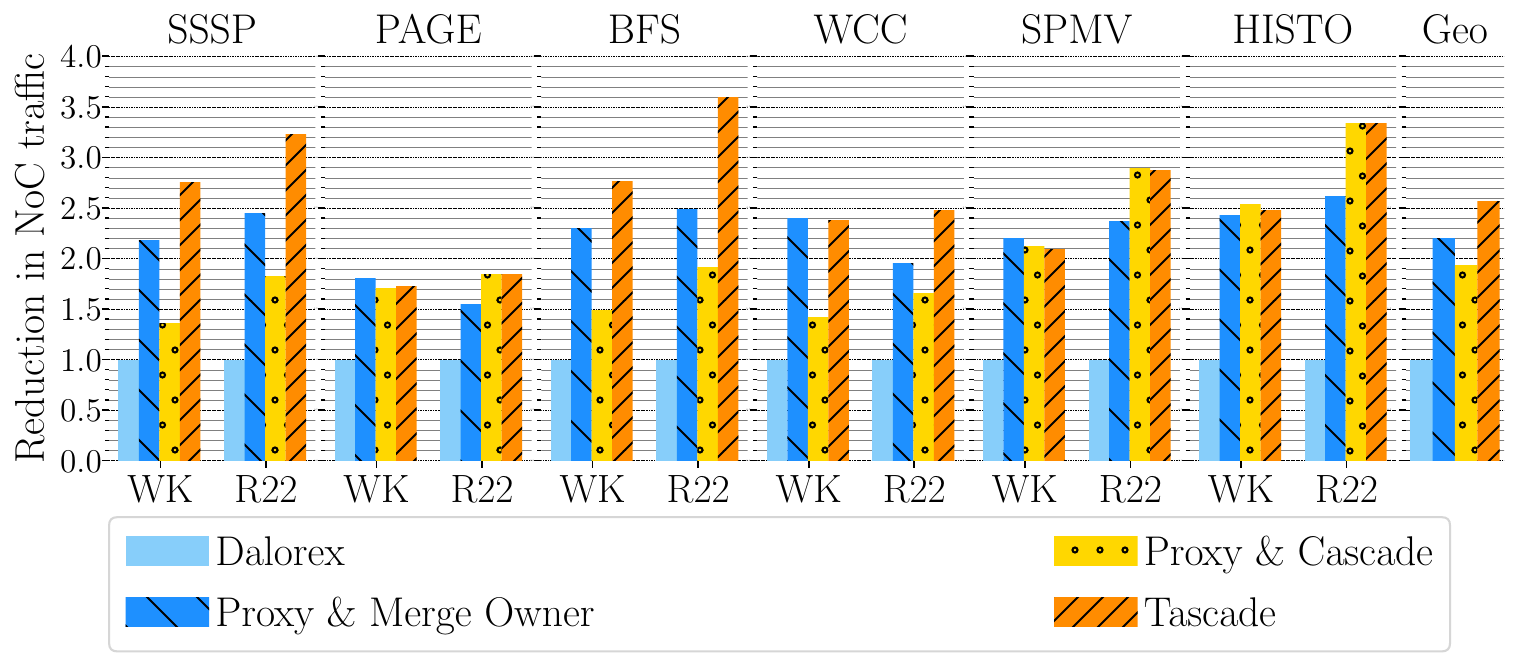}
\vspace{-5mm}
\caption{
Performance, energy efficiency and traffic-reduction gains of the accumulative features of \proj over the baseline of Dalorex (no proxy).
}
\vspace{-3mm}
\label{fig:selective_cascading}
\end{figure}

This section demonstrates the performance improvement of utilizing proxy regions, where data updates can be coalesced or filtered at proxy tiles.
\cref{fig:selective_cascading} characterizes the individual benefits of proxy regions as well as cascading, using Dalorex's $128\times128$ monolithic grid (without proxy regions) as baseline.
The proxy regions are of size $16\times16$, for which $P_{array\_local}$ fits entirely on the tile's SRAM.

\textbf{\textit{Impact of Proxy Caching:}}
\cref{fig:selective_cascading} shows that compared to the baseline, merging the proxy data directly (without cascading) into the owner tile (\textit{Proxy \& Merge Owner}) already provides a performance improvement of 4.3$\times$ geomean.
Additionally, it improves energy efficiency by 1.4$\times$, in part due to the decreased NoC traffic (2.2$\times$).
For applications operating in write-back mode (PageRank, SPMV and Histogram), the P-caches provide both coalescing as well as filtering benefit reducing overall traffic.
For applications operating in write-through mode (SSSP, BFS and WCC), updates are propagated immediately, and the main advantage of proxy comes from filtering non-minimal updates, i.e., reducing traffic.

\textbf{\textit{Impact of Cascading:}}
While proxy caching at the local proxy region alone significantly improves the performance, there are additional gains that can be achieved by continuing to do so at other proxy regions en route to the owner tile.
This cascading approach effectively implements a reduction tree across the grid (Fig. \ref{fig:reduction_tree}) where the owner tile acts as the root and the proxy tiles as the nodes of the tree.
\cref{fig:selective_cascading} dissects these improvements by evaluating cascading at every proxy tile (\textit{Proxy \& Cascade}) and selectively (\textit{Tascade}).

\textit{Proxy \& Cascade} improves performance by $1.2\times$ geomean over \textit{Proxy \& Merge}, and $5.2\times$ over the baseline.
However, its energy efficiency is 3\% worse than the baseline.
Although cascading at every region theoretically minimizes the traffic the most and ensures that all the proxy owners store the most up-to-date values, it increases the latency for the owner tile to see the updates, especially when proxy tiles are busy.
This increased latency can cause data staleness, affecting  the work efficiency of barrierless graph applications (SSSP, BFS and WCC) negatively.
Moreover, since all the proxy owners must process all cascading updates, it increases the PU energy consumption over no or selective cascading.

Selective cascading, what we call \textit{Tascade}, allows the proxy tiles en route to capture a task invocation when there is network traffic ahead or when the proxy tile is eager to process the task (i.e., low occupancy on the proxy task's IQ).
\cref{fig:selective_cascading} shows that with this selective policy, \proj improves the performance further to a $6\times$ geomean over the baseline, while also improving energy efficiency by $1.2\times$.
While \proj increases PU energy due to the extra tasks to be processed, it decreases one of the main sources of energy---the NoC---since the traffic decreases significantly, $2.6\times$ over Dalorex.

\subsection{Optimal Proxy Region Size}\label{sec:res_proxy_region_size}

The importance of the choice of proxy region size is evident when one considers the two extremes of proxy region size: a single tile and the entire grid.
On one end, with a single tile region, all tiles would have to cache an entire $P_{array}$ creating a high storage cost or low P-cache hit rate when the P-cache size is limited. In addition, cascading would be considered at every tile.
On the other end, with the proxy region size equal to the entire grid, one would recover the same configuration as Dalorex.
Therefore, we expect there to be a middle ground where the performance-optimal proxy region size lies.
Moreover, as described in \cref{sec:proxy_cache_config}, the smaller the proxy region the larger the fraction of the $P_{array}$ a tile is a proxy for ($P_{array\_local}$), and the more frequently values are evicted if we consider a constant P-cache size.

\begin{figure}[!htp]
\includegraphics[width=\columnwidth,height=2.9cm]{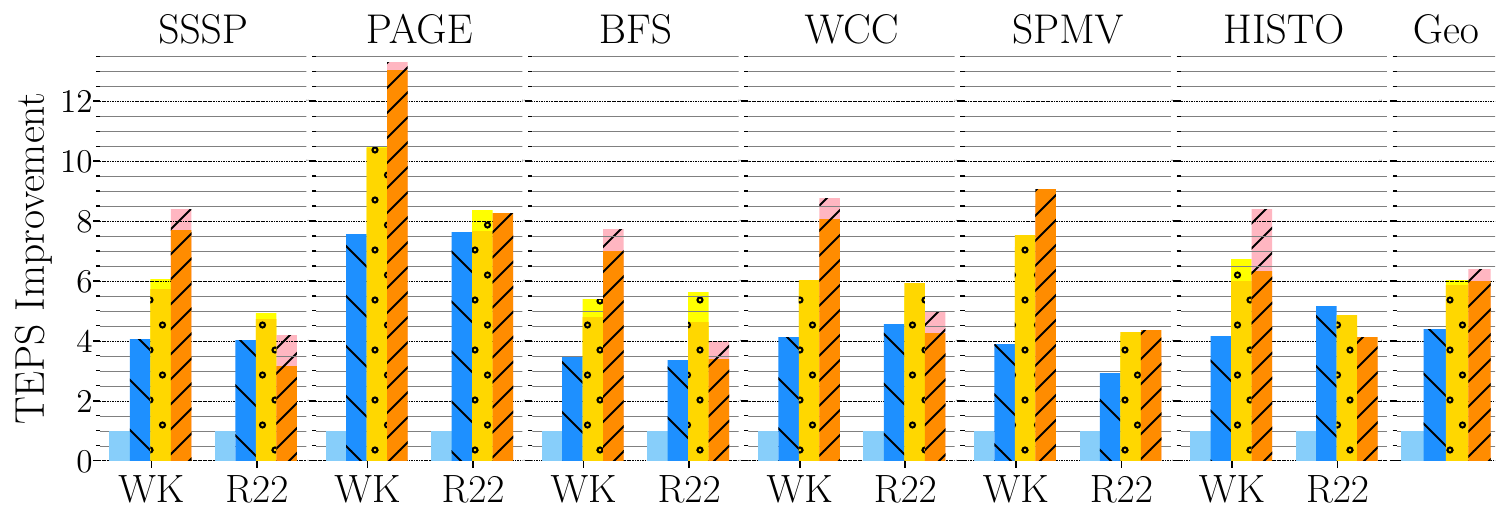}
\includegraphics[width=\columnwidth,height=2.9cm]{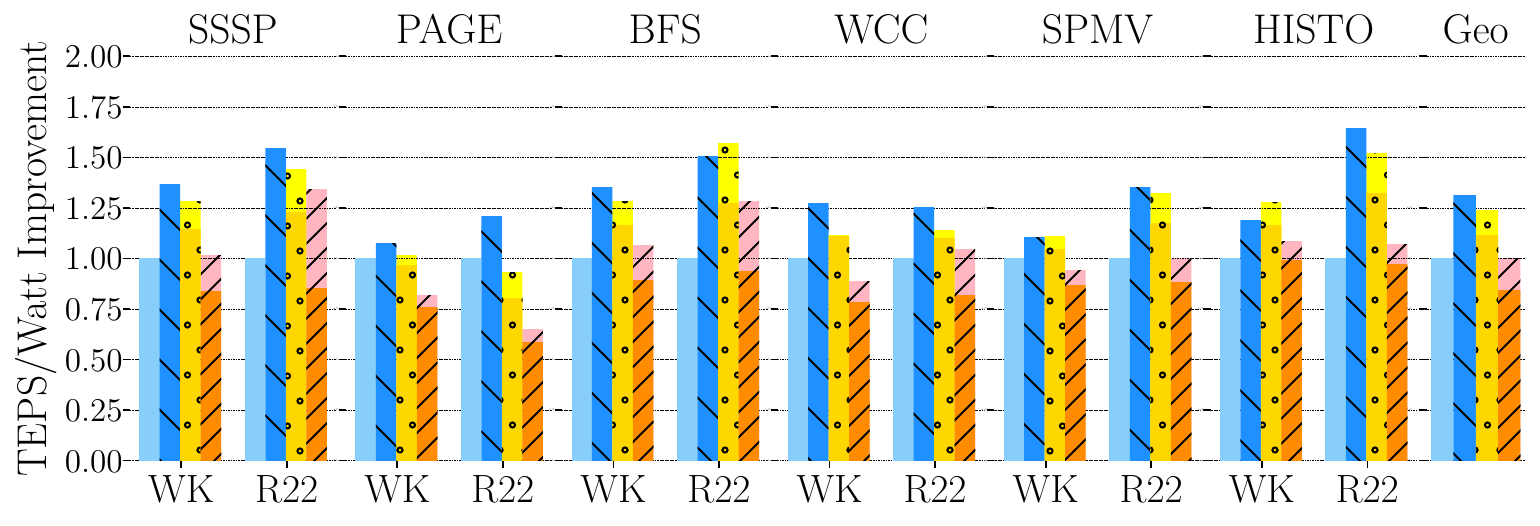}
\includegraphics[width=\columnwidth,height=3.3cm]{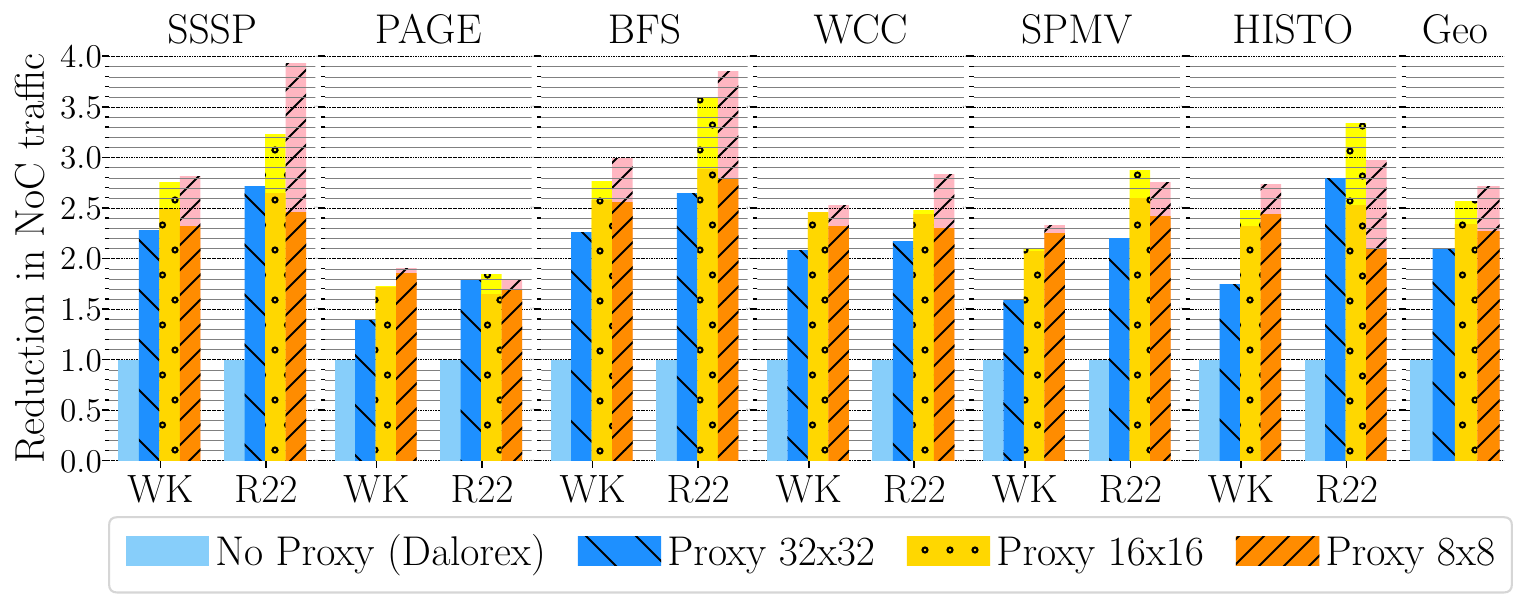}
\vspace{-5mm}
\caption{
Performance, energy efficiency and traffic-reduction gains of decreasing proxy region sizes normalized to the baseline of Dalorex.
\vspace{-4mm}
}
\label{fig:proxy_comp}
\end{figure}

For a total grid size of $128\times128$, \cref{fig:proxy_comp} shows the evaluation of the impact of proxy region sizes of $32\times32$, $16\times16$ and $8\times8$ on performance.
The bars for the last two options overlay two cases: when increasing $P_{array\_local}$ sizes are stored in the P-cache in full (light-colored bars) demonstrating peak gains and when the P-cache size is kept constant (dark-colored bars) at 16 KiB corresponding to the size of $P_{array\_local}$ for $32\times32$ regions, demonstrating the tradeoff between a smaller proxy region and a larger $P_{array\_local}$.
In the unconstrained case, performance increases as the proxy region size decreases.
However, with a limited cache size, the $8\times8$ does not significantly improve over the $16\times16$ case.
This is because with a smaller cache size values get evicted more often, leading to less coalescing of the updates.

\subsection{Impact of Limiting P-cache Capacity}\label{sec:res_proxy_cache_size}

To understand the performance impact of limited P-cache capacity, we evaluated $16\times16$ regions with decreasing SRAM budgets allocated for the P-cache.
These sizes range from 64 KiB ($P_{array\_local}$ size for this dataset and region size) to 1 KiB.\footnote{Recall from \cref{sec:proxy_cache} that the effective P-cache size is smaller than the SRAM allocated for it due to the tags.}

\begin{figure}[!htp]
\includegraphics[width=\columnwidth,height=2.9cm]{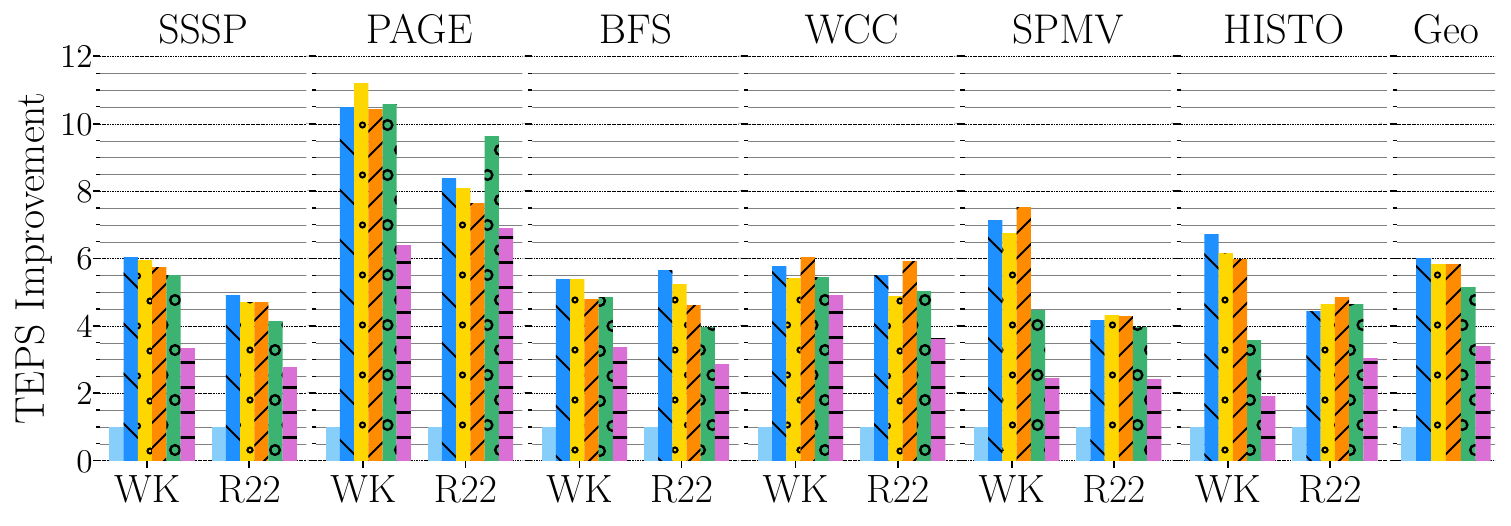}
\includegraphics[width=\columnwidth,height=2.9cm]{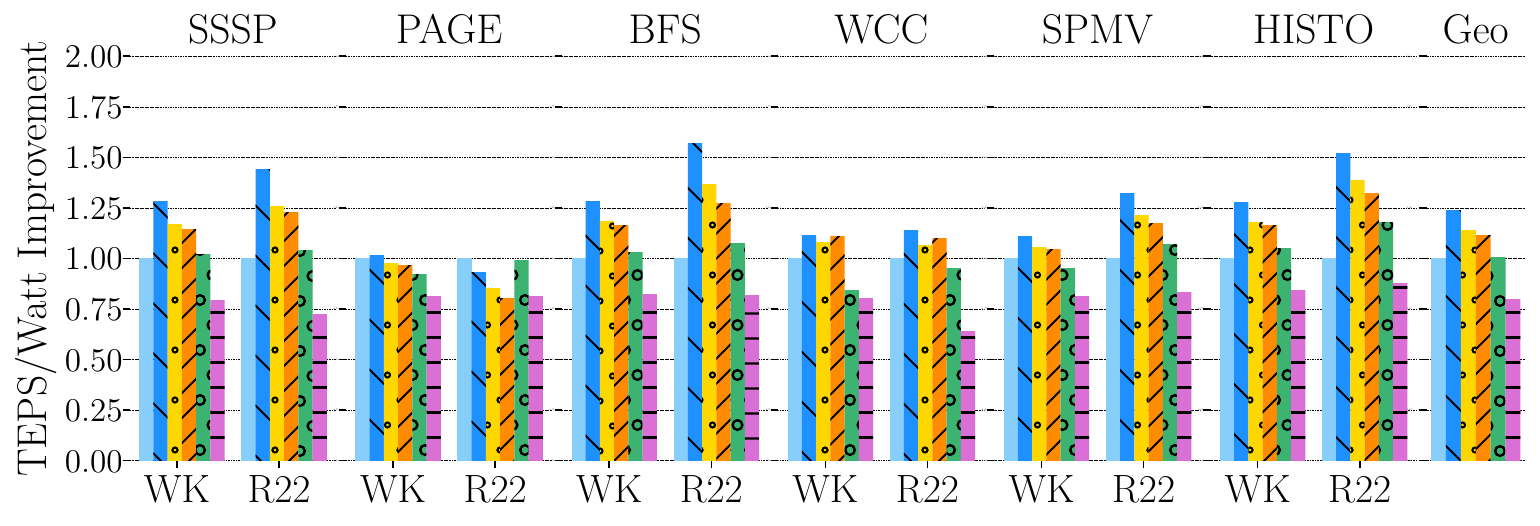}
\includegraphics[width=\columnwidth,height=3.3cm]{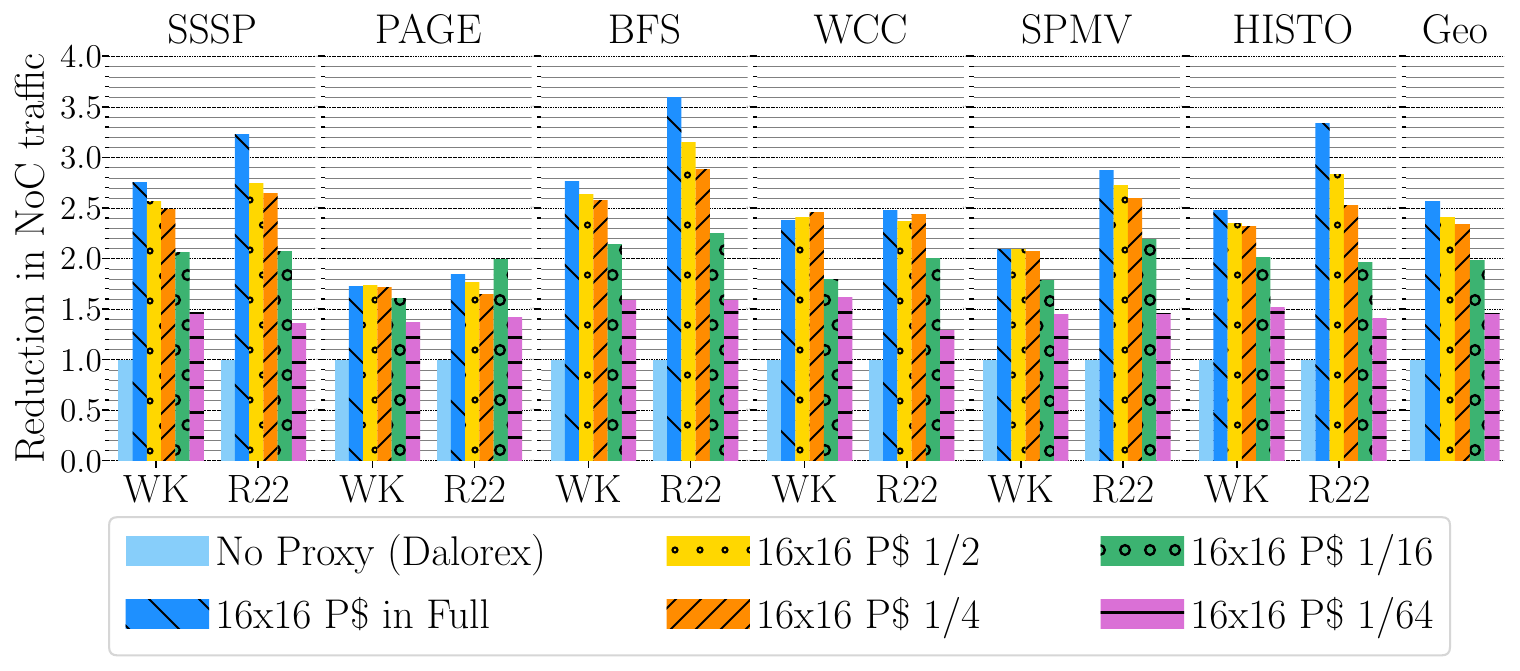}
\vspace{-5mm}
\caption{
Performance, energy efficiency and traffic-reduction gains of decreasing P-cache sizes (for $16\times16$ regions) normalized to the baseline of Dalorex (no proxy) using a $128\times128$ grid.
}
\vspace{-4mm}
\label{fig:proxy_pressure}
\end{figure}

\cref{fig:proxy_pressure} (top) shows some differences in the performance impact of P-cache size reduction across applications and datasets, but they all remain well above the baseline of Dalorex.
While in some cases performance remains stable or decreases in the first halving steps (e.g., BFS and SSSP), in others it may even increase despite the pressure on the P-cache (e.g., PageRank and SPMV).
This increase is caused by having fewer elements to flush from the P-cache towards the end of the program---since they were already merged into the owner tile upon eviction.
On geomean, the performance remains around the $6\times$ mark (over Dalorex) until the P-cache budget is reduced by $\geq 16\times$.
The performance of 1/16 and 1/64 cases (with 4 KiB and 1 KiB of storage) is still above the baseline, with a 5.2$\times$ and 3.4$\times$ improvement, respectively.

\cref{fig:proxy_pressure} also displays the gains in energy efficiency with proxy regions (middle), which are correlated with the savings in NoC traffic (bottom).
Improvement in NoC traffic over the baseline range from 2.6$\times$ to 1.5$\times$ geomean from the full P-cache size to the smallest one (1/64).
Energy efficiency is also lower with more constrained cache sizes, however, the geomean gains remain above the baseline in all but the last case.

\textbf{Takeaway:}
Proxy regions and P-caches reduce the storage overhead of reduction trees by:
(a) sharing the responsibility of a copy of the reduction array across a region of tiles, and
(b) only storing part of that copy in the P-cache, where we have seen that even a ratio of 1/16 keeps most of the benefits. 
Compared to a software-managed approach that would store a copy of the reduction array per PU, \proj decreases storage overheads by over three orders of magnitude.

\subsection{Asynchrony Improves Performance}\label{sec:asynchrony}

One of the main advantages of implementing proxy caching and selective cascading with a task-based data-local parallelization scheme is that it allows for reductions to be merged asynchronously.
This is implemented seamlessly thanks to the P-cache hardware support introduced in this paper.
Estimating the impact of asynchrony is especially important since software approaches to reduction trees often utilize such a synchronization step.
We study the cost of synchronization by evaluating the proxy approach with and without a barrier prior to merging the proxy data.

\begin{figure}[!htp]
\includegraphics[width=\columnwidth,height=2.9cm]{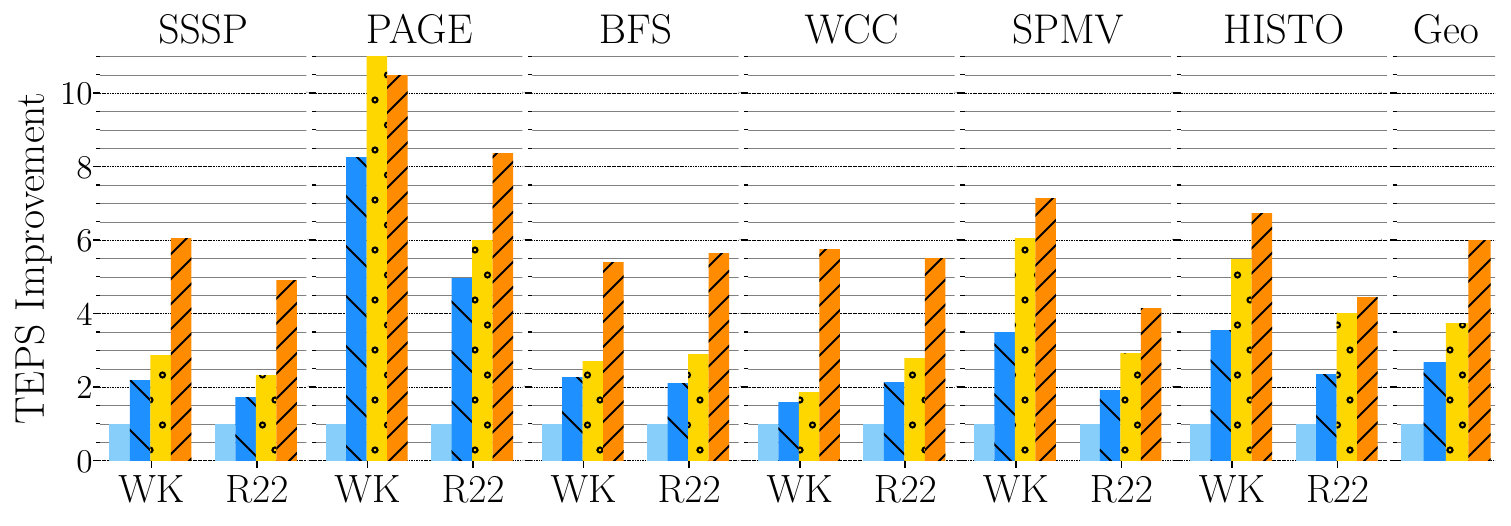}
\includegraphics[width=\columnwidth,height=3.5cm]{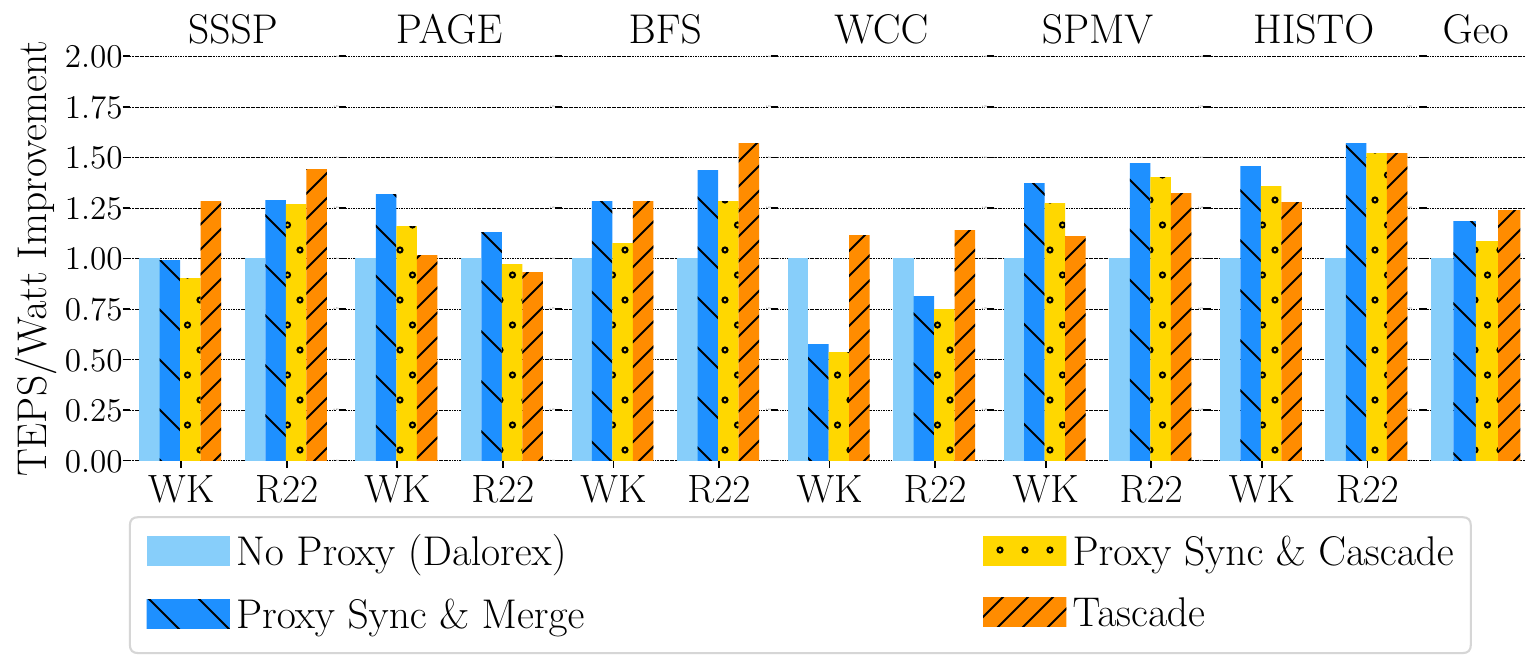}
\vspace{-5mm}
\caption{
Characterization of the performance and energy efficiency of \proj and two versions with a barrier synchronization before merging all proxy data ($16\times16$ regions) with and without cascading, normalized to the baseline of Dalorex using a $128\times128$ grid.
}
\vspace{-3mm}
\label{fig:dalorex_proxy_cascading}
\end{figure}

\cref{fig:dalorex_proxy_cascading} shows the performance of merging the proxies directly at the data owner (\emph{Sync \& Merge}) or via full cascading (\emph{Sync \& Cascade}) after the barrier is reached by all the PUs.
\proj yields a 1.6$\times$ improvement over \emph{Sync \& Cascade} and 2.3$\times$ over \emph{Sync \& Merge}, thus showing the benefits of asynchrony.

Since in this experiment, the P-cache fits the entire $P_{array\_local}$, \cref{fig:dalorex_proxy_cascading} also showcases the runtime improvement of starting to flush the P-caches towards the end of the program when PUs are often idle, instead of waiting for every PU to reach the barrier.
Moreover, merging asynchronously in \proj improves energy efficiency by 14\% over \emph{Sync \& Cascade}.

\textbf{\textit{Comparison to Software-managed Reduction Trees:}}
The synchronous versions of the proxy approach evaluated above represent an upper bound to the performance expected from a software-managed approach since the hardware components introduced by \proj provide additional benefits.
For example, these versions still use the P-cache in hardware instead of a software-managed copy of the reduction array.
Moreover, the cascade version shown in \cref{fig:dalorex_proxy_cascading} (yellow) is only synchronous before the cascading starts, and the cascading itself is asynchronous once it starts.

\subsection{Improvements Across Various NoC Designs}\label{sec:res_noc}

We envision that the hardware-enabled asynchronous reductions of \proj can be utilized in a broader set of systems ranging from server-class~\cite{amd_vcache,amd_epyc_isca,amd_rome,shapphire_rapids} and wafer-scale manycores~\cite{cerebras_fft,kumar_wafer}, to clusters of these chips connected~\cite{fugaku,fugaku_hmc}.
Since the 2D torus utilized in Dalorex is not a common NoC found in AI-oriented manycores~\cite{groq,tesla_dojo,esperanto}, we also wanted to evaluate the performance improvement \proj provides with a 2D mesh.
In addition, as an alternative to the monolithic implementation, one may use server-class-sized chips, connected with a board-level or cluster-level interconnect.
Thus, we also performed experiments evaluating the improvements of \proj when using an inter-chip interconnect as well.

\begin{figure}[!htp]
\includegraphics[width=\columnwidth,height=2.9cm]{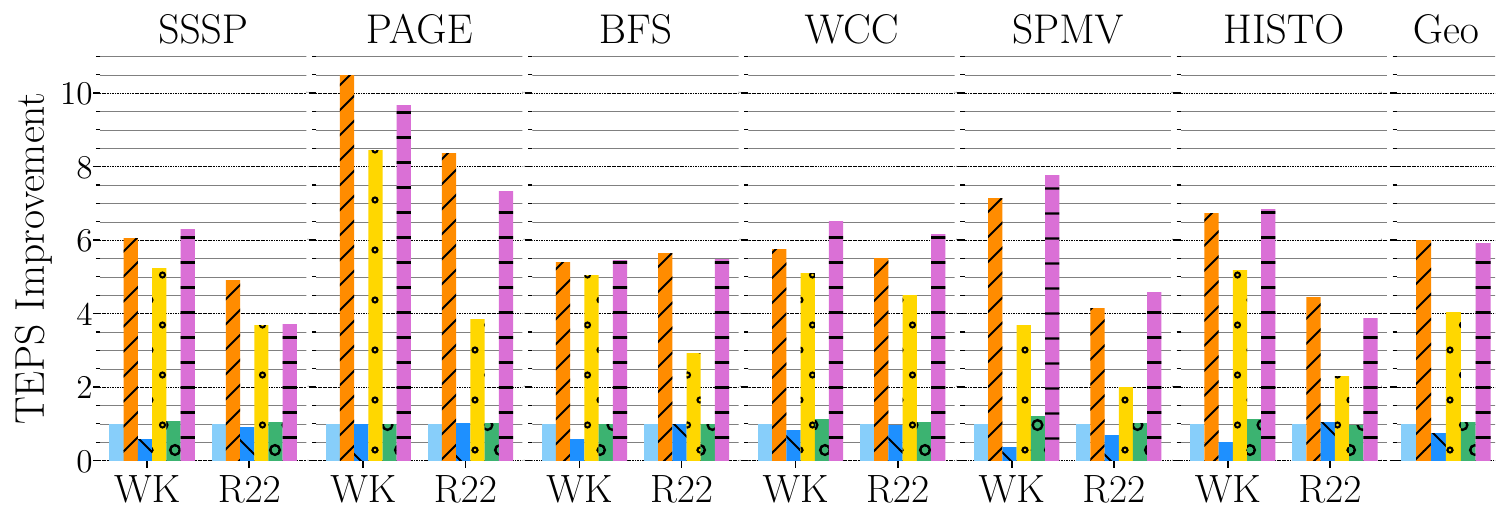}
\includegraphics[width=\columnwidth,height=3.5cm]{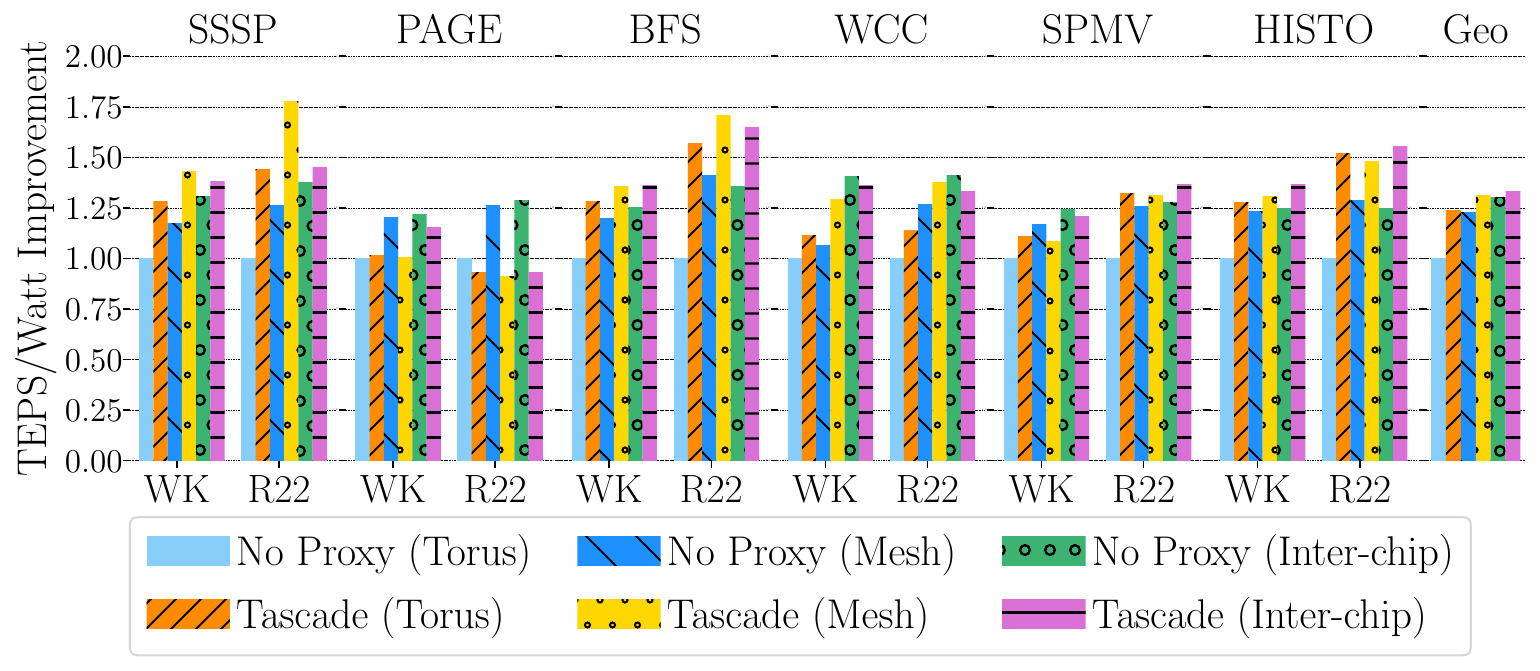}
\vspace{-5mm}
\caption{
Performance and energy efficiency gains of applying \proj to three different networks normalized to the baseline of Dalorex's torus.
}
\vspace{-3mm}
\label{fig:sync_mesh}
\end{figure}

\cref{fig:sync_mesh} shows the performance improvement of \proj with mesh, torus, and inter-chip networks, over the baseline of Dalorex (torus), for the same grid sizes ($128\times128$).
\proj yields large performance improvement over no proxy for all NoC types with 5.7$\times$, 6$\times$ and 5.4$\times$ geomean, for multi-chip, monolithic torus chip, and monolithic mesh chip, respectively and improves energy efficiency for all NoC types by $\sim25\%$.

\textit{\textbf{Higher Effective Bandwidth:}}
Proxy regions enable much of the data updates to be coalesced or filtered within the proxy region. 
When most of the communication is regional, the effective overall bisection bandwidth gets closer to the aggregated bisection bandwidth of each region.
\cref{fig:mesh_heatmap} demonstrates this with the heatmap of the PU and router activity on a mesh NoC without proxy regions (left) and utilizing Tascade (right).

\begin{figure}[!htp]
\centering
\resizebox{\columnwidth}{!}{
\begin{tabular}{c c}
\subf{\animategraphics[width=10cm, height=8cm]{5}{HEAT64M_2_64_Kron22_PU_Active/heatmap_PU_Active_frame_}{1}{38}}
{}
&
\subf{\animategraphics[width=10cm, height=8cm]{5}{HEAT8M_2_64_Kron22_PU_Active/heatmap_PU_Active_frame_}{1}{12}}
{}
\\
\subf{\animategraphics[width=10cm, height=8cm]{5}{HEAT64M_2_64_Kron22_Router_Active/heatmap_Router_Active_frame_}{1}{38}}
{}
&
\subf{\animategraphics[width=10cm, height=8cm]{5}{HEAT8M_2_64_Kron22_Router_Active/heatmap_Router_Active_frame_}{1}{12}}
{}
\end{tabular}
}
\caption{
Animation of the mesh NoC PU and router activity when running BFS on R22 in Dalorex (left) and Tascade (right).
Router activity denotes messages being routed; no activity can mean that the router has no messages to route or that the NoC is clogged and messages are stuck.
The animation is composed of snapshots at a rate of a frame per 40 microseconds, thus, the number of frames indicates runtime (38 for Dalorex and 12 for Tascade).
The animation can be visualized by opening this PDF with Adobe.
For convenience, we have also added this animation as a GIF in our \repository.
}
\vspace{-3.5mm}
\label{fig:mesh_heatmap}
\end{figure}

\textbf{Takeaway:}
As \cref{fig:mesh_heatmap} clearly visualizes, \proj:
(a)~\textit{improves the effective bandwidth of the NoC} by coalescing/filtering updates in the local proxy region and en route to the owner tile;
(b)~\textit{increases PU utilization} by allowing multiple PUs to perform data reductions without the synchronization of software-managed reduction trees;
(c)~\textit{balances NoC and PU contention} by opportunistically deciding whether proxy tasks are executed at proxy tiles (when the NoC is busy) or continue towards the owner (if the PU is busy).

\subsection{Strong Scaling Up to a Million Tiles}\label{sec:million}

\begin{figure}[!htp]
\hspace{-3mm} \includegraphics[width=1.05\columnwidth,height=3.2cm]{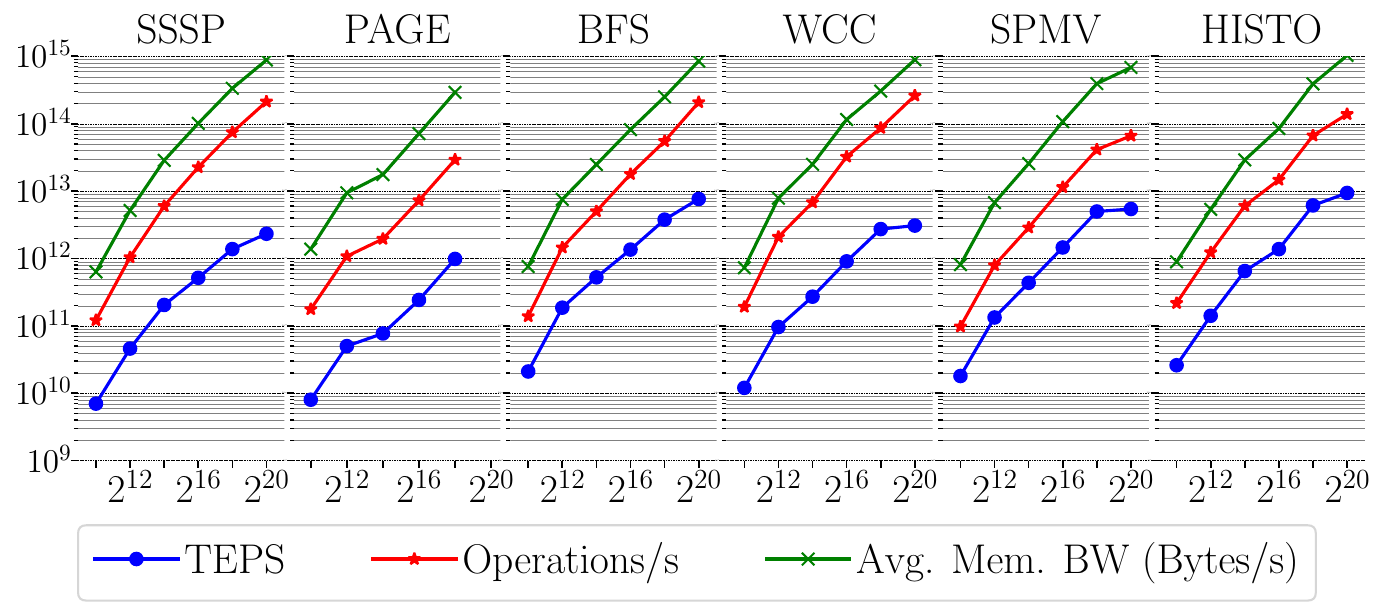}
\vspace{+1mm}
\hspace{-2mm}\includegraphics[width=1.05\columnwidth,height=3.7cm]{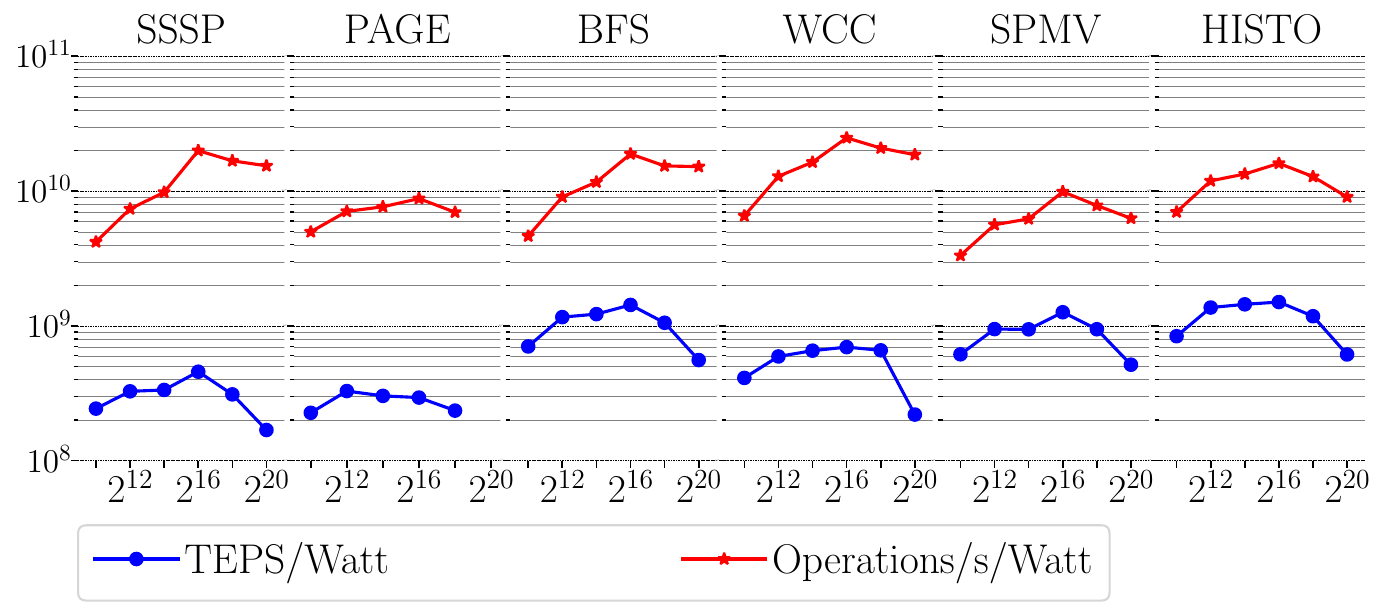}
\vspace{-5mm}
\caption{
Throughput in TEPS and operations/s, and the average on-chip memory bandwidth needed to achieve that.
The X-axis is the size of the \proj grid used when analyzing strong scaling RMAT-26, ranging from a thousand to over a million tiles.
The bottom plot shows energy efficiency.
}
\vspace{-3mm}
\label{fig:scaling_million}
\end{figure}

Up to this point, we have shown the performance improvement \proj provides over previous work and the contribution of different design components using the $128\times128$ grid ($2^{14}$ tiles) as the main frame of comparison.
\cref{fig:scaling_million} now presents absolute numbers when scaling the parallelization from a thousand PUs to over a million PUs ($2^{10}$ to $2^{20}$) by quadrupling the number at each step.
In these experiments, we use a $16\times16$ proxy region size until the $2^{16}$ parallelization, and a $32\times32$ region size beyond that.\footnote{The rationale for that is to decrease the memory footprint of the simulator itself, as the number of proxy regions scales quadratically with the grid size when keeping the proxy region size constant.}

\proj increases performance across the scaling range, achieving orders of magnitude greater scaling capability for processing sparse applications than previously demonstrated~\cite{dalorex, tesseract}.
\cref{fig:scaling_million} (top) shows how throughput scales well with the number of tiles, with some signs of plateauing at the last datapoints.
The gap between Operations/s and TEPS represents the number of instructions needed to traverse an edge (or multiply a non-zero element in SPMV).
\footnote{This efficiency gap widens with scale for barrierless graph applications (SSSP, BFS and WCC) as vertices can be re-explored before all its neighbors in the frontier have been visited or have propagated their updates~\cite{barrierless}}

\cref{fig:scaling_million} (bottom) showcases that the energy efficiency of \proj---measured by TEPS/Watt and Ops/Watt---remains fairly stable in this range of scaling, only decaying towards the last datapoints.
Note that these are extreme parallelization levels already, i.e., on the last scaling step the $2^{26}$-vertex graph is parallelized across $2^{20}$ tiles, equating to 64 vertices per tile (and 20 times as many edges).

Further improving scalability through better data placement methods and by combining it with graph partitioning~\cite{gluon,gluon_async,metis} are possible avenues for future work.

\textbf{\textit{Petabyte/s of memory bandwidth:}}
As mentioned earlier, data-structure traversal has a low arithmetic intensity.
\cref{fig:scaling_million} demonstrates how much memory bandwidth is required to maintain a high target throughput.
For the 1-million-tile configuration, SPMV reads, on average, over a PB/s from their local memories with an arithmetic intensity under 0.1 FLOPs/byte.
At peak throughput of execution, SPMV reads 2.2 PB/s to perform 100 TeraFlop/s.
This configuration uses 1024 chip packages and draws 10 KW of power on average and 29 KW at its peak, where power density stays within the tens of mW/mm$^2$---suitable for air cooling.

\textbf{\textit{In the context of the Graph500 ranking:}}
The top entry for BFS on RMAT-26 is the Tianhe Exa-node (Prototype@GraphV)~\cite{graph500}, delivering 884 GTEPS. 
For that size, \proj achieves 3540 GTEPS with $2^{18}$ PUs (256 chips) and 7630 GTEPS with $2^{20}$ (1024 chips).
The smallest dataset size with an entry higher than 7000 GTEPS is RMAT-36, which is a thousand times larger than RMAT-26.
Since weak scaling is more attainable than strong scaling (e.g., Argonne's Mira or Fugaku~\cite{fugaku}), we would expect \proj to achieve even higher throughput for datasets of this size.

For RMAT-22, the best performing prior work demonstrates up to 70 GTEPS~\cite{simula_graphcore_bfs} running these codes~\cite{liu2015enterprise,wang2016gunrock} on a V100-SXM3 GPU.
For that size, our $2^{16}$ PU configuration---evaluated in \cref{fig:dalorex_comparison}---yields 1,760 GTEPS (25$\times$ higher).

\vspace{-1mm}
\section{Conclusion}\label{sec:conclusions}

This paper introduces lightweight hardware support for scalable reductions on manycore systems.
The novel integration of the P-cache coupled with software-reconfigurable region sizes, enables efficient coalescing and filtering of reduction operations.
This is further enhanced by the opportunistic and asynchronous propagation of data updates, leveraging our router's selective cascading and P-cache's write-propagation policy.
The P-cache minimizes the storage needed for reduction trees and does not introduce additional hardware storage, as it efficiently utilizes the existing SRAM per tile. 

The evaluation of \proj underlines its effectiveness by demonstrating substantial performance improvements across various network topologies and scales.
We characterize the accumulated benefits of software-configurable proxy regions, proxy caches, and selective cascading, over prior work Dalorex, where improvements increase with scale, ranging from $6\times$ geomean for 16K PUs to $14\times$ for 64K PUs, across several graph and irregular applications.
\proj scales performance up to 1 million PUs for a billion-edge graph, achieving 7630 GTEPS for BFS, which in the context of the Graph500 list, it is 8.6$\times$ higher than the top entry for that graph size.
These results not only validate the technical contributions of this paper but also establish \proj as a highly scalable and efficient solution for graph processing.

\section*{Acknowledgments}
\noindent
This material is based on research sponsored by the Air Force Research Laboratory (AFRL), Defense Advanced Research Projects Agency (DARPA) under agreement FA8650-18-2-7862, and National Science Foundation (NSF)
award No. 1763838.~\footnote{The U.S. Government is authorized to reproduce and distribute reprints for Governmental purposes notwithstanding any copyright notation thereon. The views and conclusions contained herein are those of the authors and should not be interpreted as necessarily representing the official policies or endorsements, either expressed or implied, of NSF, AFRL and DARPA or the U.S. Government.}

\bibliographystyle{IEEEtranS}
\bibliography{refs}

\end{document}